# The Deeper, Wider, Faster Program: Exploring stellar flare activity with deep, fast cadenced DECam imaging via machine learning

S. Webb[1,2]⋆, C. Flynn[1,2], J. Cooke[1,2,3], J. Zhang[1,2], A. Mahabal[4] T.M.C. Abbott[5], R. Allen[1], I. Andreoni[4], S.A. Bird[6,7,8], S. Goode[1,2], M. Lochner[9,10], T. Pritchard[11]

[1] *Centre for Astrophysics and Supercomputing, Swinburne University of Technology,*
*Mail Number H29, PO Box 218, 31122, Hawthorn, VIC, Australia*
[2] *ARC Centre of Excellence for Gravitational Wave Discovery (OzGrav), Australia*
[3] *ARC Centre of Excellence for All Sky Astrophysics in 3D (Astro3D), Australia*
[4] *Division of Physics, Mathematics and Astronomy, California Institute of Technology, Pasadena, CA 91125, USA*
[5] *NOIRLab, Mid-Scale Observatories/Cerro Tololo Inter-American Observatory, Casilla 603, La Serena, Chile*
[6] *China Three Gorges University, Yichang 443002, People's Republic of China*
[7] *Center for Astronomy and Space Sciences, China Three Gorges University, Yichang 443002, People's Republic of China*
[7] *Key Laboratory of Optical Astronomy, National Astronomical Observatories, Chinese Academy of Sciences, 20A Datun Road,*
*Beijing 100101, People's Republic of China*
[9] *African Institute of Mathematical Sciences, Muizenberg, Cape Town, 7950*
[10] *South African Radio Astronomical Observatory, Observatory, Cape Town, 7295*
[11] *Center for Cosmology and Particle Physics, New York University, New York, NY 10003*

## ABSTRACT

We present our 500 pc distance-limited study of stellar flares using the Dark Energy Camera as part of the Deeper, Wider, Faster Program. The data was collected via continuous 20-second cadence $g$ band imaging and we identify 19,914 sources with precise distances from Gaia DR2 within twelve, ~3 square-degree, fields over a range of Galactic latitudes. An average of ~74 minutes is spent on each field per visit. All light curves were accessed through a novel unsupervised machine learning technique designed for anomaly detection. We identify 96 flare events occurring across 80 stars, the majority of which are M dwarfs. Integrated flare energies range from ~ $10^{31}$ – $10^{37}$ erg, with a proportional relationship existing between increased flare energy with increased distance from the Galactic plane, representative of stellar age leading to declining yet more energetic flare events. In agreement with previous studies we observe an increase in flaring fraction from M0 – M6 spectral types. Furthermore, we find a decrease in the flaring fraction of stars as vertical distance from the galactic plane is increased, with a steep decline present around ~100 pc. We find that ~70% of identified flares occur on short timescales of < 8 minutes. Finally we present our associated flare rates, finding a volumetric rate of 2.9 ± 0.3 × $10^{-6}$ flares $pc^{-3}$ $hr^{-1}$.

**Key words:** Parallaxes, Stars: Flare, Stars: Low-mass, Methods: data analysis

## 1 INTRODUCTION

Stellar flares occur stochastically across main sequence stars, commonly seen in those with with large convective envelopes, and result from violent reconnection of magnetic field lines (Haisch et al. 1991; Pettersen 1989; Lacy et al. 1976). This sudden release of magnetic energy into the sur-rounding stellar atmosphere causes rapid dissipation of chromospheric and coronal field and sheet aligned currents, generating strong emission seen in the radio through to X-ray, with the largest emission occurring within the UV continuum (Osten et al. 2005; Hawley & Pettersen 1991; Lin & Hudson 1976; Berger et al. 2008). To date, stellar flares have been found to occur across a vast range of energies, spanning from nanoflares of $10^{23}$ erg recorded from our own Sun by Parnell & Jupp (2000), to superflares releasing upwards of ~ $10^{33}$–$10^{36}$ erg as discovered by Maehara et al. (2012)

⋆ E-mail: swebb@swin.edu.au

within the data from *The Kepler Space Telescope* (Borucki et al. 2010, *Kepler* hereafter). The flare frequency distribution as a function of integrated flare energy is a power law with higher energy flares being less likely (Pettersen 1989; Ishida et al. 1991; Shibayama et al. 2013; Chang et al. 2015).

Bopp & Moffett (1973) first identified two apparent populations of flares, classical and complex, based on their temporal structure, dividing them by number of peaks present, with classical flares only exhibiting one peak and complex flares having multiple peaks (see also Moffett 1974). Classification may be limited by survey cadence, where low temporal resolution of the flare event is observed. More recently Dal & Evren (2010) further identified sub-populations of fast and slow flares from the ratio of rise to decay time, with ratios of < 3.5 identified as slow and > 3.5 as fast. Work by Hawley et al. (2014) showed for M-dwarfs, complex flares not only have longer durations on average but also dominate the larger energy ranges seen in their sample from *Kepler*. This correlation between complex events and longer decay times suggests the possibility of there only being two populations, being classical/fast and complex/slow. The historical classification into four populations may be due to past survey cadences and target selections. However the fast cadenced, continual imaging required to definitively test this correlation has only been achieved by few studies to date, with majority targeting specific objects for other science goals.

Stellar flares are most commonly generated by low mass stars. These cooler stars have larger convective envelopes surrounding their radiative cores, with theoretical models showing a star's structure transitioning into a fully convective regime at a stellar mass of $\sim 0.3 M_\odot$ (Dorman et al. 1989; Chabrier & Baraffe 1997; Morales et al. 2009). This effect plays a part in the observed M4 transition, where stellar activity increases dramatically between M3–M5 stellar types. Stellar activity from these stars produces flares with hot (blue) thermal components, providing a photometrically dramatic contrast against their cool red photosphere (West & Hawley 2008; Hilton et al. 2010; Pineda et al. 2013). This dramatic contrast in thermal energy allows for easier detection, in comparison to flares of similar energy releases from hotter (bluer) stars.

Several studies into flare properties have taken advantage of M dwarf properties to further understand flare activity as a function of age, duty cycles, flare rate as functions of Galactic latitude, rotation period, and light-curve shape (Hawley & Pettersen 1991; Kowalski et al. 2009; Hilton et al. 2010; Pineda et al. 2013; Maehara et al. 2012; Schmidt 2018; Günther et al. 2019; Davenport et al. 2012; Davenport 2016)). Notably, work by Kowalski et al. (2009) examined ~50,000 light curves from 10,497 M-dwarfs, collected via the Sloan Digital Sky Survey (SDSS) of Stripe 82. Kowalski et al. (2009) found a strong relationship between flaring stars and those stars which have strong Hα emission during quiescence, with the large majority of flare events occurring on late type (and younger) M-dwarfs of subtype M4-M6. Kowalski et al. (2009) also confirmed that there was a trend towards higher energetic flares occurring on the relatively inactive higher mass stars, M0-M1. These results were verified by Hilton et al. (2010) using the SDSS low-mass star spectroscopic sample of ~38,000 M dwarfs from West et al. (2008) to identify a further 63 flares. Duty cycles of these flares show a large increase with spectral subtype, increasing from

0.02% to 3% for early and late M-dwarfs respectively. The SDSS photometric and spectroscopic samples were also used to explore the relationship between flare rate and Galactic latitude, both finding that nearly all flares in their respective samples occurred near the Galactic plane, indicating the younger, later type M-dwarfs, are more predominant within stellar populations of the plane (Kowalski et al. 2009; Hilton et al. 2010). Later work on SDSS DR7 M-dwarf magnetic activity by Pineda et al. (2013) further confirmed that the active flare fraction decreases with greater radial distances from the Galactic centre due to stellar age.

The space-based *Kepler* survey has provided the largest flare catalogs to date, primarily exploring flare activity across solar-like stars during the original mission (Walkowicz et al. 2011; Hawley et al. 2014; Davenport 2016; Van Doorsselaere et al. 2017; Yang & Liu 2019). While the modified observing strategy implemented during K2 (the 2nd Kepler mission due to the failure of reaction mechanisms) allowed for the study of 134 and 540 bright M dwarfs by (Stelzer et al. 2016) and Yang et al. (2017) respectively. More recently, with the launch of the Transiting Exoplanet Survey Satellite (TESS; Ricker et al. 2009), Günther et al. (2019) uncovered a further 673 flaring M dwarfs via the 2-minute cadence data. Several ground based surveys, including the Next Generation Transit Survey (NGTS; Wheatley et al. 2018), EVRYSCOPE (Law et al. 2014), All-Sky Automated Survey for Supernovae (ASAS-SN; Shappee et al. 2012) have also produced detailed flare catalogs for bright nearby stars (Howard et al. 2018; Jackman et al. 2018; Dillon et al. 2020; Rodríguez Martínez et al. 2020a). The recent work by Chang et al. (2020) investigates M dwarf flare activity across the southern sky using the SkyMapper Southern Survey DR1. Chang et al. (2020) find 254 flare events and a steep decline in flaring fraction towards larger vertical distances from the Galactic plane.

The Deeper, Wider, Faster program (DWF; Cooke et al. in prep) commenced in 2014 and has targeted several fields with deep, fast 20-second cadenced optical imaging, which provides a unique opportunity to study flares over temporally resolved time frames. Here we present the findings of a 500 pc distance limited flare study across 12 DWF fields. Section 2 describes the DWF program and optical data collection. Section 3 outlines our analysis and methodology for flare identification and the calculation of flare characteristics. The results and discussion are presented in Section 4, and conclusions are provided in Section 5.

## 2 DATA

We describe the DWF program, the targeted fields and the criteria used when choosing them in Section 2.2, and the nature of our fast cadenced, deep optical imaging in Section 2.3.

### 2.1 The Deeper, Wider, Faster Program

Several classes of fast (millisecond-to-hours duration) optical transient events have been discovered over the last few decades and the progenitors and physical mechanisms behind many of them are still relatively poorly known (e.g., supernovae shock breakouts and other rapidly evolving extragalatic events, see Garnavich et al. 2016; Prentice et al. 2018; Perley et al. 2018). What has limited our ability to detect and understand these events has been the capability to gather deep data in short, regular time intervals before, during and after the events as well as over a range of wavelengths, i.e., deep, wide-field, fast-cadenced non-targeted surveys. The DWF has been designed with these challenges specifically in mind, constructing a multi-wavelength and simultaneous observational program of over 80 facilities to date[1]. DWF takes a 'proactive' approach to transient astronomy, with multi-wavelength observations of the target fields taken continuously over 1-3 hour periods, typically over 6 consecutive days, acquiring data before, during and after transient events. From our real time processing we are able to rapidly identify candidates and coordinate rapid and long-term follow-up observations. DWF was first created in 2014 and since its inception has had two commissioning runs and eight operational runs (see Andreoni & Cooke (2018), Cooke et al., in prep).

| Target Field | RA | DEC | Galactic Latitude | Total Time on Field (hr) | Dates Observed | # Sources[a] (⩽**500 pc**) |
|---|---|---|---|---|---|---|
| Antlia | 10:30:00.0 | −35:20:00.0 | 19.172 | 7.02 | 3,5-7 Feb 2017 | 1743 |
| Dusty10 | 10:12:00.0 | −80:50:00.0 | −19.957 | 3.50 | 7,9 June 2018 | 1911 |
| Dusty12 | 11:46:00.0 | −84:33:00.0 | −21.889 | 4.39 | 26-30 July 2016 | 2091 |
| FRB131104 | 06:44:00.0 | −51:16:00.0 | −21.930 | 6.37 | 14-17 Jan 2015, 18-22 Dec 2015 | 1776 |
| 8hr | 08:16:00.0 | −78:45:00.0 | −22.618 | 4.95 | 6-9 June 2018 | 1805 |
| Dusty11 | 11:20:00.0 | −85:20:00.0 | −22.814 | 2.35 | 7-9 June 2018 | 1790 |
| NGC6744 | 19:08:00.0 | −64:30:00.0 | −26.054 | 18.38 | 26, 28-30 July 2016, 2-7 July 2016 | 1966 |
| Prime | 05:55:07.0 | −61:21:00.0 | −30.262 | 6.88 | 14-17 Jan 2015, 2-7 Feb 2017 | 1459 |
| NSF2 | 21:28:00.0 | −66:48:00.0 | −39.823 | 3.64 | 26-27 July 2016 | 1453 |
| FRB010724 | 01:18:06.0 | −75:12:19.0 | −41.804 | 2.63 | 18-22 Dec 2015 | 1436 |
| 4hr | 04:10:00.0 | −55:00:00.0 | −44.756 | 8.05 | 15-17 Jan 2015, 18-22 Dec 2015 | 1252 |
| 3hr | 03:00:00.0 | −55:25:00.0 | −53.432 | 6.88 | 18-22 Dec 2015 | 1242 |

**Table 1.** The 12 DWF fields used in this work. Total time on field indicates the combined observations over the several dates observed..
[a] Average number of sources over all observation epochs, with ⩾3 $\sigma$ detections.

The unique design of DWF allows exploration of transients on milliseconds-to-hours timescales, providing further understanding into the classes of already observed fast transient events as well as exploring events theorised to occur on these timescales. The optical component of DWF is able to explore a region of parameter space not yet reached by previous ground based transient surveys, by taking continuous, high cadenced 20 second exposures, imaging with the wide-field sensitive Dark Energy Camera (DECam, FOV=3 deg$^2$ Flaugher et al. 2015) on the 4m Blanco telescope in Chile, or continuous 30 second exposures using the Hyper Suprime Cam (HSC, FOV=1.8 deg$^2$ Furusawa et al. 2018; Komiyama et al. 2018; Kawanomoto et al. 2018; Miyazaki et al. 2018) on the 8m Subaru telescope in Hawaii. Note: this work will only focus on the data gathered from DECam. Here, we present findings from our in-depth analysis of stellar flares discovered in the post-run archival DECam data processing.

### 2.2 DWF fields used

In this work we use a total of 12 DWF fields, each visited multiple times between the years 2015–2018, as shown in Table 1. The total time visited on each field varied, averaging 6.30 hours collected over several nights. A total of 75.08 hours of observations are analysed in this work.

The ongoing DWF program has targeted over 20 distinct fields, with several having repeat visits. Due to the nature of DWF, targeted fields need to have simultaneous visibility by Chilean facilities (DECam, Gemini, VLT, etc), as well as Australian or South African radio telescopes (ASKAP, Parkes, Molonglo, MWA, MeerKAT), Antarctic telescopes (South Pole Telescope, AST3-2), and space-based facilities, as well as the growing number of other simultaneous and follow-up facilities (more detail can be found in Cooke et al., in prep). The fields are selected using several criteria, mainly, (1) sky position, enabling common visibility for the multiwavelength telescopes to simultaneously observe during runs scheduled at different times of the year, (2) preference towards low-redshift galaxy clusters/groupings or globular clusters, (3) legacy fields with multiwavelength pre-imaging, and (4) fields with previously detected Fast Radio Bursts (FRBs) and FRB repeaters, as well as the need for regions of low Galactic extinction for certain wavelength facilities. As a result, the fields targeted are somewhat arbitrarily distributed across the sky, providing data across a variety of Galactic declinations.

### 2.3 DECam fast-cadence Imaging

The imaging data were collected over multiple DWF runs using DECam. During a DWF run, continuous 20-second $g$ band exposures are taken across multiple target fields. The use of the $g$ band filter maximizes the depth of our imaging, reaching ~0.5 magnitudes deeper in comparison to the other filters in dark time. The expected limiting magnitude in $g$ band is m(AB) ~ 23, for an average seeing of 1.0 arc-seconds and airmass of 1.5 (relatively high airmass due to the field constraints of observing simultaneously with multiple facilities). All images used in this work were post-processed through the NOAO High-Performance Pipeline System (Valdes & Swaters 2007; Swaters & Valdes 2007;

---

[1] http://bit.do/DWF

Scott et al. 2007), and transferred to the OzSTAR super-computer at Swinburne University of Technology. We processed each of the 62 CCDs in parallel, performing source extraction via SExtractor (Bertin & Arnouts 1996), zero-point corrections and magnitude offset corrections against the SkyMapper DR2 catalog (Bertin & Arnouts 2010; Onken et al. 2019). A master list was compiled by cross-matching the position of all extracted sources from each CCD, over all exposures within an 0.5 arcsec radius between source centroids, producing one catalog of unique source positions per field. We used this master catalog for reference when selecting our volume limited sample, as explained in detail in the following section.

### 2.4  Volume Limited Star Selection

In this paper we explore flare activity out to 500 pc across each of our 12 fields covering a total sky area of ~36 square degrees and volume of ~443,000 pc$^3$. Distances to the stars were provided by the *Gaia* DR2 parallaxes to pre-select our sample (Luri et al. 2018; Evans et al. 2018). For each field, we queried *Gaia* DR2 using Vizier[2], selecting only sources that meet the following criteria:

- 1000/Parallax(mas) ≤ 500 pc
- Δ Parallax/Parallax (mas) ≤ 0.2
- G Magnitude ≥ 11

It should be noted that due to the systematic implicit biases associated with measuring parallaxes our sample of stars are more likely to be on the higher end of their associated distance ± error (as described by Trumpler & Weaver 1953). We choose to apply a magnitude limit cut to avoid sources likely to saturate from our DECam observations. Each *Gaia* source catalog was cross-matched against each field's master source catalog, with light curve files generated for each source existing in our DWF data. Finally, only light curves with 3 or more consecutive detections during a single night's visit were evaluated for flares (possibly excluding flares shorter than 40 seconds). Table 1 outlines each of our target fields, and the number of *Gaia* selected sources. A total of 19,914 sources were assessed. Separate light curves were generated for each source over each night of observations. The total number of light curves in the sample is 114,958.

## 3   ANALYSIS

### 3.1  Flare Identification

#### 3.1.1  Anomaly Ranking Visual Inspection - candidate flagging

To identify flares within our sample, we first use unsupervised machine leaning to rank light curves from most to least anomalous, as detailed in full in Webb et al. (2020). A set of 25 features was extracted for each light curve in our sample. The features and light curves were then fed into the python based *Astronomaly* package (Lochner & Bassett 2020). *Astronomaly* consists of a python back end and JavaScript front end to easily explore the data via a locally hosted web interface. We pass the light curve features through the scikit-learn implementation of isolation forest within *Astronomaly* to determine the most anomalous sources (Ting et al. 2008). We visually inspect all light curves through the *Astronomaly* GUI, shown in order of anomaly ranking (as measured by the isolation forest algorithm). All light curves with potential flares are flagged for further evaluation. We reduce the original 114,958 light curves to just ~700 during this process by removing static stars using HDBSCAN clustering (for details see Webb et al. 2020).

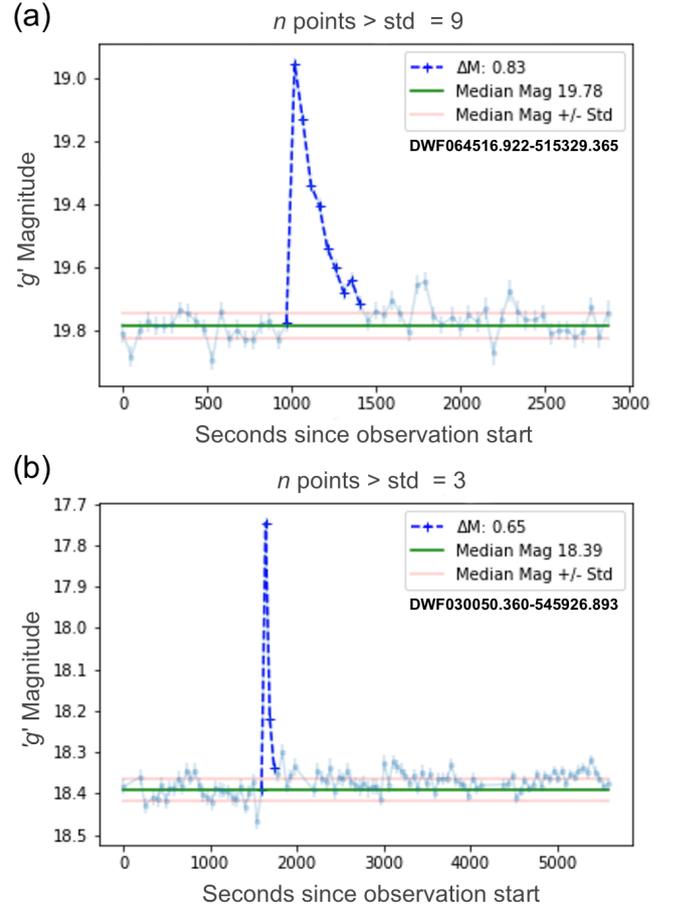

**Figure 1.** Classical and short-duration flare examples. Panel (a) An example flare with 9 consecutive data points greater then one standard deviation (red line) above the median magnitude (green line) of the quiescent source. Panel (b) An example short-duration flare with 3 consecutive data points greater then one standard deviation (red line) above the median magnitude (green line) of the quiescent source.

#### 3.1.2  Flare detection and criteria

The flagged light curves were then passed through a Change Point Analysis (CPA) algorithm to flag the start and finish positions of individual flares within the light curve. We use the method as first outlined by Chang et al. (2015) to develop a flexible CPA algorithm, modified to suit our fast cadenced data. CPA is useful in identifying abrupt variations in sequential data and is done so by combining a cumula-tive sum scheme and bootstrap rank statistics to flag points of systematic change within sequential data. Our adaption of Chang et al. (2015) FINDflare algorithm is used to flag all possible flares within our shortlisted light curves. The criteria used within the FINDflare algorithm are as follows, N1: The required number of standard deviations above the median for the data points, N2: The number of standard deviation plus uncertainty above the median, N3: The number of consecutive points required to meet the above criteria to be flagged as a flare. We choose to use the values N1=3, N2=2, N3=3, after sensitivity testing on known DWF flares identified from previous work by Andreoni et al. (2020) and Webb et al. (2020). We then estimate the quiescent magnitude of each source by taking the median magnitude of all data points before the flare (and in very rare cases of flares at the beginning of the observations, after the flare). We then determine the standard deviation, $\sigma$, of the quiescent section of the light curve, and as set with the N3 value, require at least three consecutive data points to be greater then the two standard deviations above the median magnitude, see Figure 1 for example. All light curves which passed these criteria were manually inspected alongside their image cutouts for each exposure. This stage of careful image evaluation allowed the rejection of candidates which where caused due to observational effects (e.g. cosmic rays, hot pixels, edge detections).

### 3.2 Flare Characteristics

For each flare in our sample we record several characteristics of the event. These include, duration, maximum change in magnitude ($\Delta M$), flare type and Equivalent duration..

The duration of the flare is measured from the last observed quiescent point. The first data point, of 2 consecutive data points, to fall below the upper 1 sigma uncertainty to the mean quiescent magnitude, two examples can be seen in panel (a) of Figure 2. In the rare event of our observations finishing before the flare reaches the star's quiescent magnitude, the duration is a minimum limit and marked accordingly in Table A and an example can be seen in panel (b) of Figure 2.

The $\Delta M$ of each flare is calculated by the difference between the peak magnitude recorded and the median quiescent magnitude for each source. Here it is important to note that these measurements can underestimate the true value of $\Delta M$ which could occur during the 20–30 second readout after each exposure and diluted by the 20 second exposure itself. This is more likely for the short duration flares. Figure 1 demonstrates both longer duration (a) and short duration flares (b). These characteristics are recorded further in Table A.

### 3.3 Integrated Flare Energy

The integrated flare energy represents the total amount of energy released over the duration of the flare. To calculate the total integrated flare energy we first calculate the g band component from our observations.

To estimate total energy in $g$ band, each light curve was normalised by setting the star's quiescent flux to one, creating a relative output for each data point. Using the

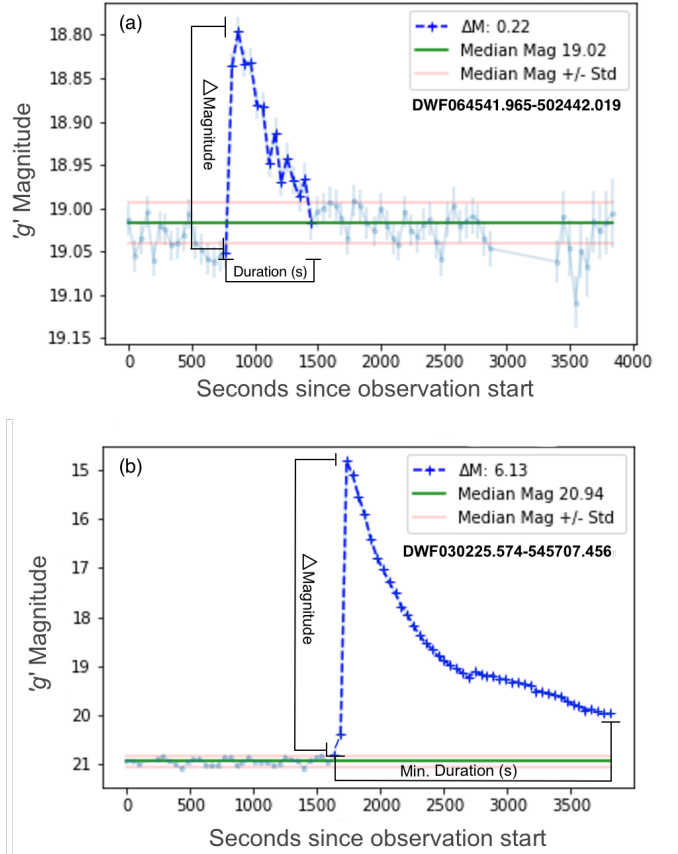

**Figure 2.** Panel (a): Flare star example illustrating the maximum magnitude change and total event duration definitions. The maximum change in magnitude is determined by the difference between the peak flare detection and the median quiescent magnitude (green line; see definition of the quiescent mag described in section 3.1.2). The total duration of the event is measured from the last detection of a quiescence magnitude before the flare, to the point at which the flare re-enters within 1 $\sigma$ of the mean quiescent magnitude range (red lines). Panel (b) a flare event example with the flare continuing past the observational period, permitting only a minimum duration to be recorded.

relative outputs over the duration of the flare, we sum the total excess amount of flux generated by the flare event, $\sum_{rel}$. Figure 3 demonstrates a flare's excess flux for each point (relative intensity >1). We use the Riemann sum binning method across all points. For each flare we first derived the relative intensity for each point in the light curve from the magnitudes with the following:

$$\frac{I_{0+f}}{I_0} = 10^{\frac{\Delta mag}{2.5}} \tag{1}$$

where $I_{0+f}$ and $I_0$ are the intensity values of the flaring and the quiescent stellar surfaces in the observed $g$ band. Note, this method may underestimate the true local flux by a small amount small spikes in flux after reaching the quiescent level threshold are not included. For each detection of the flare, the relative intensity is integrated over the flare duration to

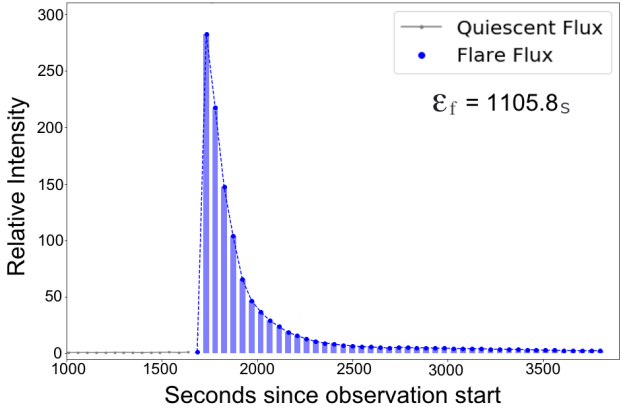

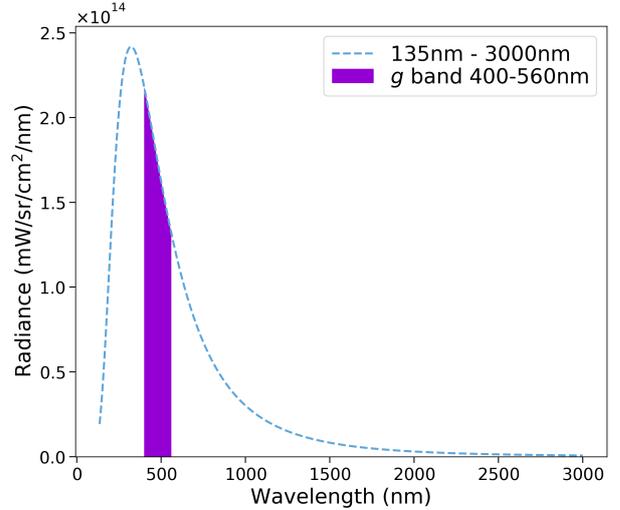

**Figure 4.** Planck function of spectral radiance of a 9000 K object. The *g* band emission component (400-560nm) is highlighted purple. Note that not all emission from across the 400-560nm is included in the ratio calculation, as it was corrected for relative transmission of flux in 50nm blocks from the DECam *g* filter. Function generated using RADIS python package.

calculate the Equivalent Duration:

$$\varepsilon_f = \int_{t_1}^{t_2} \left( \frac{I_{o+f}(t)}{I_o} - 1 \right) dt. \tag{2}$$

The quiescent stellar fluxes for *g* band, $F_\lambda$, are computed taking the median quiescent magnitude of the source, and computing the flux $(W/m^2)$ for the specific wavelength of 430 nm via the open source Gemini Observatory conversion tool[3].
The flux at source is then calculated with the following:

$$F_* = (1000 \times F_\lambda) \times (4\pi d^2) \tag{3}$$

where *d* is distance to star,. Finally the total integrated *g* band flare energy is calculated by multiplying the relative flare energy by the quiescent stellar flux and the duration of the flare:

$$E_{f(gband)} = \varepsilon_f F_* t \tag{4}$$

where t is duration of the flare in seconds.

To estimate the total integrated flux across all wavelengths of the flare we modeled the expected blackbody emission for a typical flare. Flare temperatures have been found to vary dramatically across similar spectral types and energy ranges. Commonly, the combined line and continuum emission of flares has been approximated by a ~9,000 K blackbody (Osten & Wolk 2015). However recent work by Howard et al. (2020) found that across 42 K5–K5 superflares, 43% of flares emit temperatures above 14,000 K. In this work will assume a likely flare temperature of 9000 K.

To calculate the overall contribution of the *g* band flux emission, we produce a Planck function for *T* = 9000*K* and compute the ratio of radiance over the DECam *g*-band width (400–560nm), taking into account the relative transmission across the filter, and compare it to the full emission, see Figure 4. From this we find that the total emission across our *g* band accounts for 10.9% of the total energy released. We therefore multiply the $E_{f(gband)}$ by 9.2 to account for the full bolometric energy $(E_{f(bol)})$.

## 4   RESULTS AND DISCUSSION

In our sample of 19,914 stars, we identified a total of 96 flares, from 80 individual stars, within our DWF 500 pc distance limited sample. We breakdown our flare results across source spectral type, Galactic latitude, stellar age and flare duration. We further explore the Flare Frequency Distribution (FFD) of our sample, and present our flare rate calculations.

### 4.1   Spectral Types

Using Gaia G magnitudes and BP-RP colour information, we identified the likely spectral type for each source in our sample, identifying that this work primarily explores G – M6 spectral types. Flaring sources were identified across the K to M6 range, with the majority occurring across the M2–M5 range (Figure 5). Figure 6 (a) shows the number of stars as a function of spectral type in our 500 pc sample and those which flared, and (b) presents the flaring fraction for each of the binned spectral types. We find an increase in flaring fraction across the M dwarfs, with it peaking in our sample at ~ 14% for M6. Although the overall flaring fractions are moderately lower then previous work (~ 30% lower for M5), we still observe a steep increase from M4 onward, coinciding with where it is thought M dwarfs become fully convective (Kowalski et al. 2009; Yang et al. 2017; Mondrik et al. 2019; Günther et al. 2019; Rodríguez Martínez et al. 2020b).

Our lower flaring fractions may be due to limitations on continuous time on target fields, with an average of ~74 minutes on field per night. We speculate that we are unable to observe the full range of classical longer duration flares (e.g. Hawley et al. 2014; Günther et al. 2019), through our observation strategy and selection criteria (e.g. the flare peak must be within the data). As discussed in more detail in Section 4.5, our flare sample predominately contains shorter duration events.

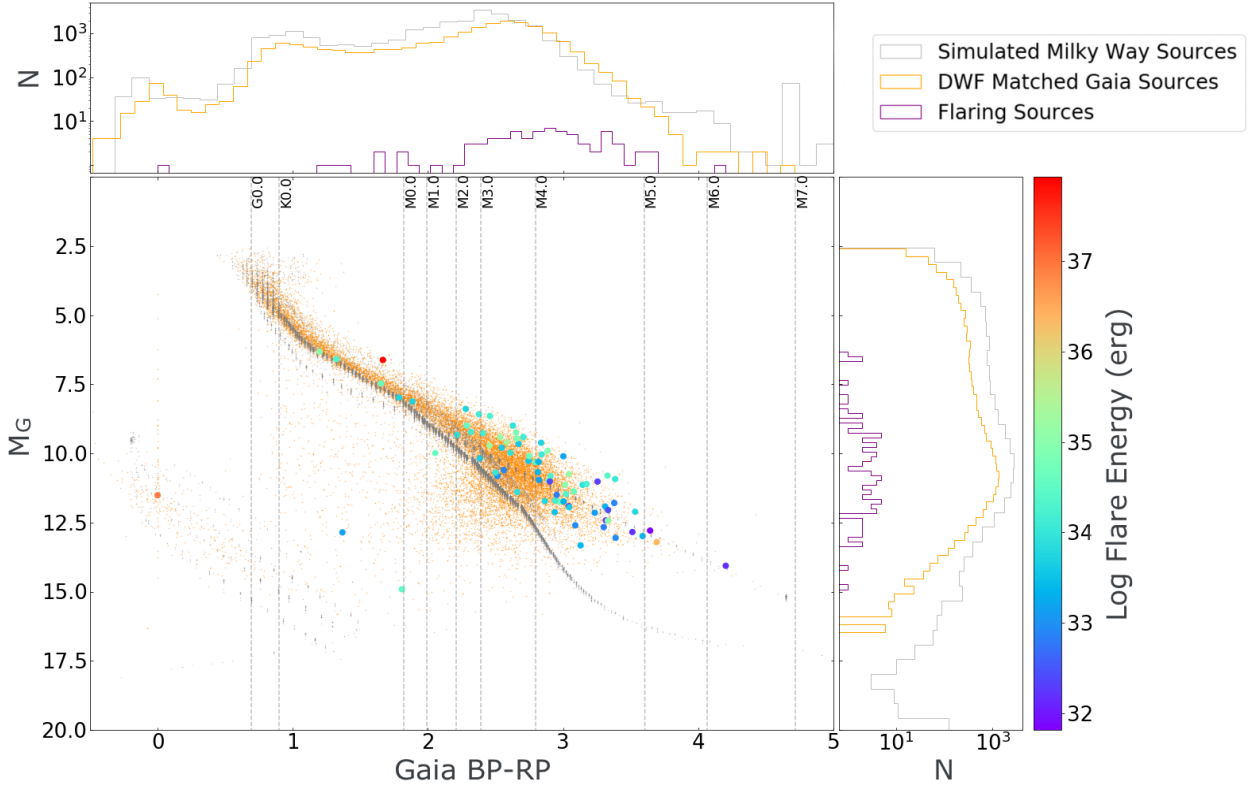

**Figure 5.** Colour-Magnitude diagram representing all sources within this study (orange ⊙), simulated sources for this study's fields (black +) and the identified flaring sources (●). The simulated sources were retrieved from Vizier's online catalog of the Milky Way stars from the Gaia Universe Model Snapshot (GUMS). The histograms for both the x and y axis show in more detail the distributions of sources across our sample, simulated sample and specifically flares (Robin et al. 2012).

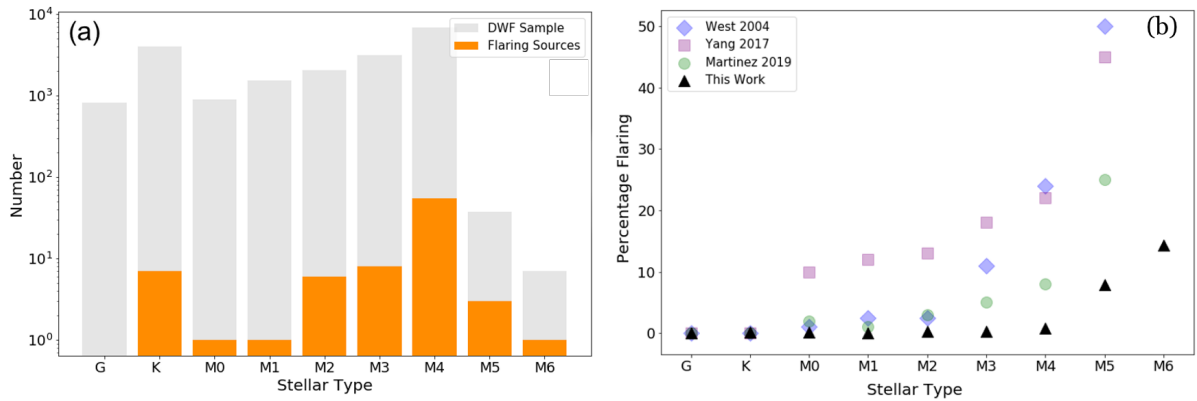

**Figure 6.** *Panel a):* Spectral type distribution for full DWF sample and the flaring sources. The spectral types are derived from the intrinsic colour in V − I from Ducati et al. (2001) by applying the transformation GBP − GRP = (V − I) + 0.15, as outlined in Mould et al. (2019). *Panel b):* Percentage of flaring sources versus spectral type, with the spectral type determined in the same manner as the previous panel. Values as identified by West et al. (2004), Yang et al. (2017) and Rodríguez Martínez et al. (2020b) are also plotted as diamonds, squares and circles, respectively.

Using the likely spectral types from Figure 5, we investigated the colour-colour temperature relationship previously presented in (Schmidt et al. 2016). We use the All WISE (Cutri et al. 2021) W1 (3.35um) and W2 (4.6um) bands to calculate (W1 − W2). To calculate (*r-z*) we used the Skymapper DR3 (Wolf et al. 2018; Onken et al. 2019) and *z* band observations for each of our flaring sources. Both (W1 − W2) and (*r-z*) values for sources can be found in Appendix A. The resulting colour–colour plot can be seen in Figure 7. We use the BP − RP colour temperature relation-ship as outlined by Casagrande et al. (2020) to estimate the effective temperature ($T_{eff}$) for each of our flaring sources, coloured accordingly on Figure 7. For each spectral type we calculate the median (W1 − W2) and ($r-z$) positions and $T_{eff}$, plotted in green markers. The estimates for $T_{eff}$ breakdown for redder sources and are not reliable for sources with (BP − RP) > 2.5. Those spectral types which have untrustworthy temperature estimates are presented by red markers. We expand the initial temperature relationship presented by (Schmidt et al. 2016), probing to the M3 range, however are unable to present reliable temperature colour relationships for M4–M6 range.

### 4.2  Flare Energies

Figure 8 shows flare duration versus integrated flare energy for our sample of flares alongside additional flare samples from various surveys. We identify a population of low energy flares, bridging the gap between X class solar flares ($\sim 10^{31}$ ergs) and the majority of stellar flares as cataloged by Kepler and GALEX ($\sim 10^{35}$ ergs). These low energy flares have a small range in duration, occurring over $\sim$ 5–30 minute periods. Interpreting the scaling relations in Figure 8 indicates our sample is especially diverse, spanning magnetic field strengths of 60 to several hundred Gauss and active region length scales of $10^9$ cm to exceeding the radius of the Sun.

Interestingly, Figure 8 also demonstrates a clear relationship between distance above the Galactic plane $Z$ and total flare energy. Our sample appears to show flares further from the Galactic plane have larger total integrated flare energy. Total flare energy also tends to be larger for longer duration flares.

We find that the distribution of flare energies is bimodal as shown in Figure 9, with this grouping of lower energy flares being centralised around sources $\lesssim 100$ pc from the Galactic plane. To ensure that this result was not due to observational selection effects, we took the detected flares with distances < 100pc and simulated them at increasing distances out to 500 pc. To do this we incrementally added 100 pc to the source distances, creating four samples in the following ranges, 100-200 pc, 200-300 pc, 300-400 pc, 400-500 pc. To calculate the magnitudes of each source at each simulated distance we used the following,

$$m = 5 \times log_{10}(d + x/d) + q_{mag} \quad (5)$$

where $m$ is the magnitude at the simulated distance, $d$ is the original distance of the flaring source, $x$ is the additional distance added (e.g. 100 pc, 200 pc etc), and $q_{mag}$ is the original quiescent magnitude of the flaring source. The difference between $m$ and $q_{mag}$ is calculated and subsequently applied for each point in the original light curve and newly associated magnitude errors were calculated. The magnitude errors were modeled using a DECam magnitude error versus magnitude plot (for a typical night of average seeing and airmass see Appendix B1). The simulated sample was put through the identical flare identification pipeline, described in Section 3.1.

We find 95% of our <100 pc flare sample were recovered, when simulated between 100-200 pc distances. The recovery rate continues to drop out to 500 pc, reaching 60%, see Figure 10. If we were assume the stellar population is isotropic in our sample, we would expect to see a ∼30–40% more flares at these distances then our study observes, suggesting a drop off of low energy flares due to the young disc. Via these tests we also confirm that several lower energy flares, $\sim 10^{31}$ergs were detectable at 500 pc. However the drop in number of low energy flares past ∼100 pc in our sample suggests an astrophysical cause, likely stellar age, further discussed in Section 4.3.

### 4.3  Flares across Galactic declination

We postulate that the results discussed Section 4.2, of densely populated flares within ∼100 pc, is representative of the dense population of stars within the thin disc of the Milky Way. The thin disc has an estimated scale height of 200–300 pc (Cabrera-Lavers et al. 2007; Jurić et al. 2008; Abazajian et al. 2009) with the youngest stars concentrated closest to the plane (Binney & Tremaine 1987). More specifically it is the younger M-dwarfs which are predominant within the stellar population of the thin disc (Kowalski et al. 2009; Hilton et al. 2010; Pineda et al. 2013).

To investigate the flaring fraction as a function of the vertical distance from the Galactic plane $Z$ we use the distance, as calculated using the Gaia DR2 parallaxes (Luri et al. 2018), and each field's central Galactic latitude to determine the vertical distance from the Galactic plane. Across our 12 fields, we probe an average distance of ∼247 pc off the Galactic plane. Table 2 displays the maximum $Z$ for each field, and the average $Z$ for flares identified in each field. On average the majority of flares are occurring at ∼ 1/3 of the maximum $Z$ probed.

When combining data across all fields, we find a strong relationship between flaring fraction and $Z$ (normalised over the total amount of observing hours), shown in Figure 11. This result is in good agreement with previous studies (e.g. Kowalski et al. 2009; Brasseur et al. 2019; Chang et al. 2020), that also find this relationship.

### 4.4  Flares from young stellar sources

Stellar flare activity is known to decrease over stellar lifetimes. This occurs as a star loses angular momentum, via stellar winds, and consequently results in a quieting of the internal magnetic dynamo from decreased rotational velocity. This age-activity relationship has previously been directly connected with flare activity (e.g. Skumanich 1972; Wright et al. 2011; Davenport et al. 2019). Feinstein et al. (2020) studied young K5-M5 stars, and found that flare rates and amplitudes decreased for very young stars ($t_{age}$ >50 Myr) across all temperatures $T_{eff} \geq 4000$ K.

We expect to find a broad distribution with higher average transverse velocity $V_T$ across our full 500 pc sample, typical of the well studies age-velocity relationship within the Milky Way (Seabroke & Gilmore 2007; Rix & Bovy 2013; Mackereth et al. 2019). However, for the flare star sample of stars with low $Z$ positions, we expect these sources will have lower average $V_T$ velocity, associated with a younger age. To investigate this age-activity relationship within our sample, we utilised Gaia DR2 proper motions to calculate the transverse velocity ($V_T$), in km/s, of the stars in our

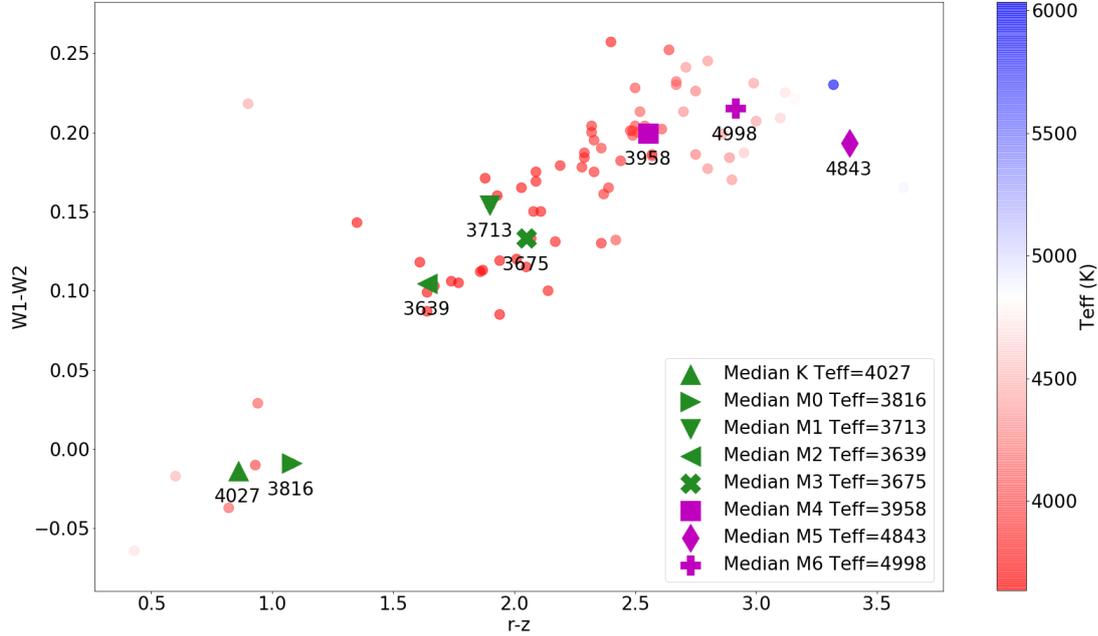

**Figure 7.** Colour-colour diagram of (W1 − W2) vs (r − z) for DWF flares plotted with circles mapped to $T_{\text{eff}}$. $T_{\text{eff}}$ was derived for each source using the Gaia (BP − RP) colour to temperature relationship outlined in Casagrande et al. (2020). The median positions for each spectral class is plotted with green markers. The colour-temperature relationship breaks down for sources with (BP − RP) > 2.5 and those spectral classes effected have their median position markers coloured in red. The temperature relationship presented by Schmidt et al. (2016) is shown with black lines and labelled.

| Field | Galactic Latitude | Max \|Z\| at 500 pc distance cut | # of Flares | Average \|Z\| of flares |
|-------|-------------------|----------------------------------|-------------|-------------------------|
| Antlia | 19.172 | 164.12 | 8 | 73.76 |
| Dusty10 | −19.957 | 170.57 | 4 | 52.47 |
| Dusty12 | −21.889 | 186.31 | 3 | 72.73 |
| FRB131104 | −21.93 | 186.64 | 13 | 65.50 |
| 8hr | −22.618 | 192.2 | 6 | 121.15 |
| Dusty11 | −22.814 | 193.77 | 1 | 63.73 |
| ngc6744 | −26.054 | 219.50 | 27 | 81.30 |
| Prime | −30.262 | 251.86 | 16 | 111.67 |
| NSF2 | −39.823 | 320.07 | 3 | 194.78 |
| FRB010724 | −41.804 | 333.14 | 6 | 36.93 |
| 4hr | −44.756 | 351.9 | 10 | 109.96 |
| 3hr | −53.432 | 401.43 | 9 | 92.96 |

**Table 2.** DWF fields galactic latitudes, maximum distance above the Galactic plane (|Z|) probed via our 500 pc distance sample, and the average |Z| of flares observed in that field

sample using,

$$V_T = 4.74 \times (\text{pmRA}^2 + \text{pmDEC}^2)^{0.5} \times D \qquad (6)$$

where pmRA is the proper motion measured in the right ascension (mas/yr), pmDEC is the proper motion measured in declination (mas/yr), and D is the distance to the source (kpc).

Using $V_T$, we can use the relationship between age and the observed transverse velocity of stellar sources within the Galactic thin disk.

To explore this we calculated the cumulative distributions of transverse velocities across our full 500 pc sample of sources, and flaring sources, shown in Figure 12. We find a median $V_T$ of 35 km/s across our full 500 pc sample of sources, and a median $V_T$ of 24 km/s for the flaring sources only. When comparing the two cumulative distribu-tions via the Kolmogorov–Smirnov test, we find a *p*-value of <1 %, suggesting two distinct distributions. Using the Gaia Universe Model Snapshot (GUMS) (Robin et al. 2012), we model the cumulative distribution of a full simulated 500 pc sample across our 12 fields and confirm that the distributions between the simulated and DWF full 500 pc are nearly indistinguishable. We also present the cumulative distribution of isolated young sources (<0.65 Gyr) from within the GUMS sample, which suggests a similar distribution to that of our flares.

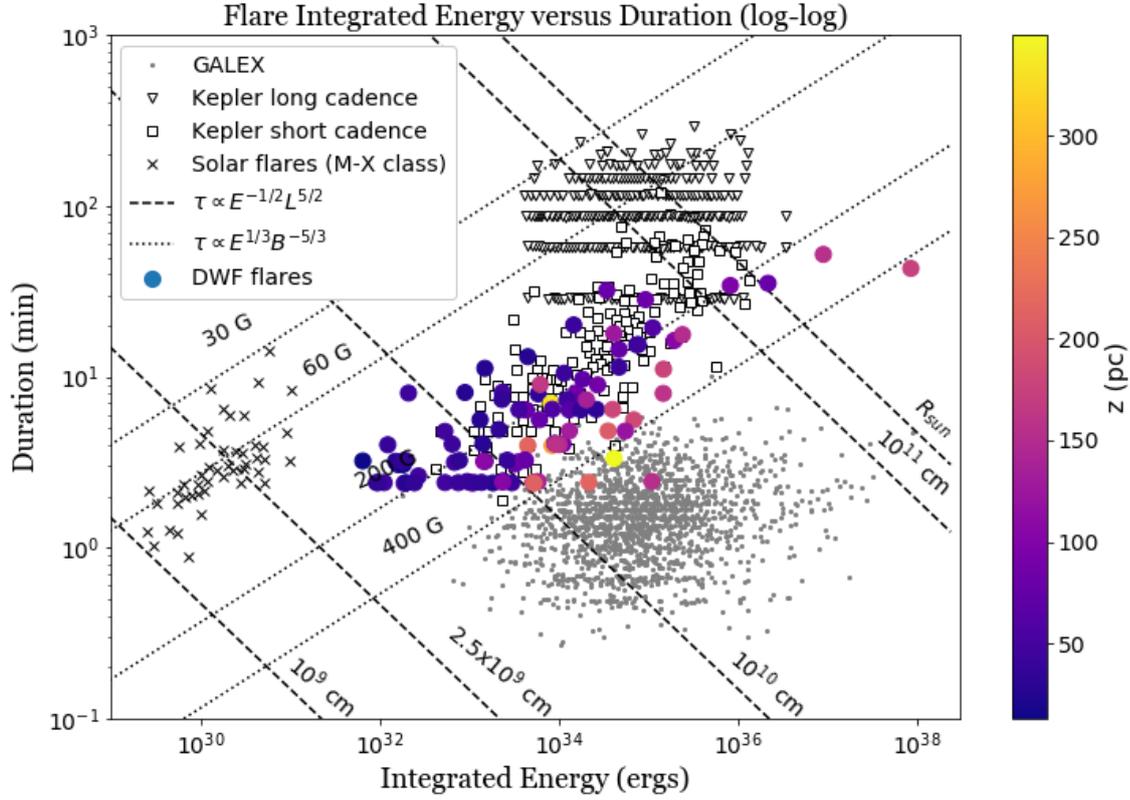

**Figure 8.** Duration versus flare energy for the DWF sample (●) with colour representing absolute distance |Z| from Galactic plane. Additionally plotted are white light solar and stellar flare datasets from Namekata et al. (2017) (×), *Kepler* superflares identified in long cadence data by Shibayama et al. (2013) (▽), Kepler superflares identified in short cadence data by Maehara et al. (2015) (□), and GALEX short duration flares identified by Brasseur et al. (2019) (·). The dotted and dashed lines show the theoretical scaling laws derived in Namekata et al. (2017), where B is magnetic field strength in the flaring region, and L is the flare length scale.

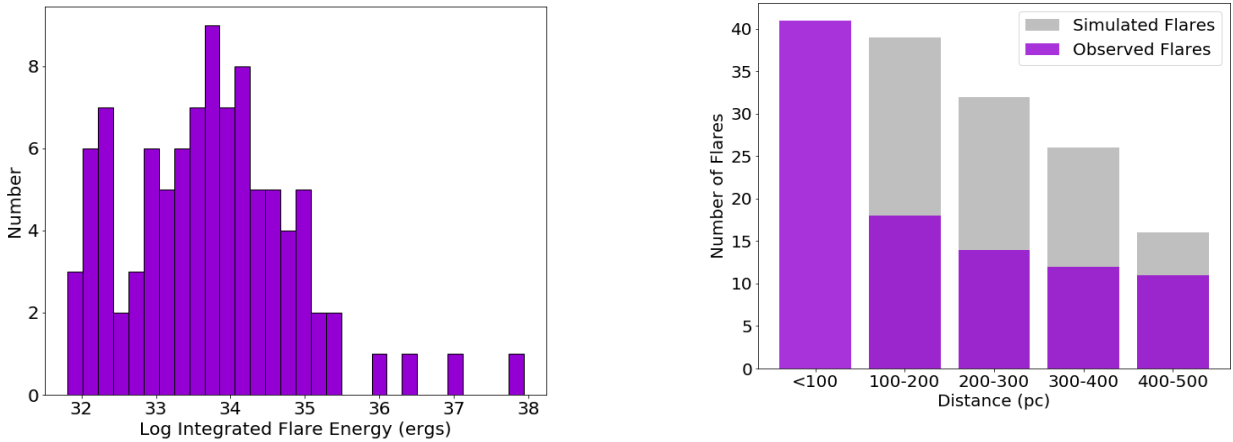

**Figure 9.** Distribution of log integrated flare energies, ranging from ~$10^{31-37}$ ergs.

**Figure 10.** Number of real flares (purple) detected in our sample grouped in 100 pc bins. The 41 flares within 0–100 pc were simulated in 100 pc bins out to 500 pc and those recovered in the flare identification pipeline are represented in grey.

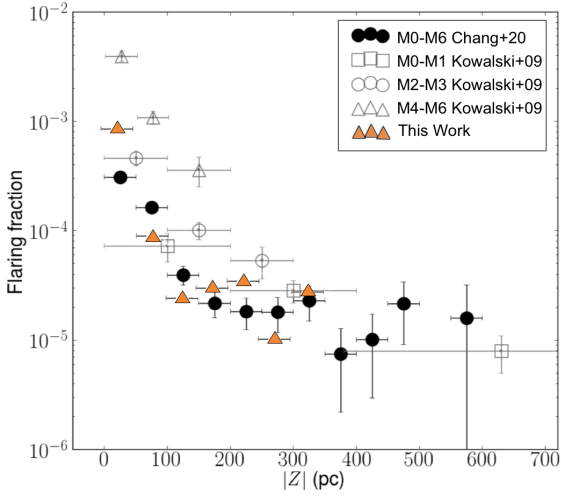

**Figure 11.** Fraction of flaring stars (normalised by total number of hours observed in this study) as a function of vertical distance from the Galactic plane |Z| (orange triangles, |Z| bin size 50 pc). We overlay results from Kowalski et al. (2009) for their spectral type ranges (M0-M1: open squares, M2-M3: open circles, M4-M6: open triangles) and from Chang et al. (2020) for their sample (M0-M6: closed circles).

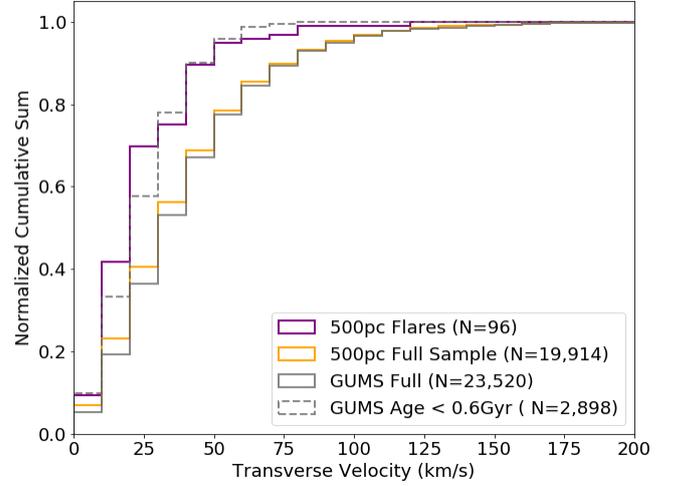

**Figure 12.** Cumulative distribution of transverse velocities for our 500 pc Flare sources (solid purple line), 500 pc full sample of sources (solid orange line), GUMS 500 pc sample (solid grey line), and 500 pc GUMS young sources (<0.65 Gyr).

From these results we conclude that the majority of flaring sources within our sample are from young stellar sources within the thin disk.

### 4.5 Short Duration Flare Population

We find a logrithmic distribution with a large duration range for our sample, as shown in panel (a) of Figure 13. Specifically, we identify that the majority of flares in our sample occur on very short timescales ($\lesssim 8$ minutes). Note that the longest flare duration continuous light curve is 156 minutes on DWF field NGC6744. The long flare durations are limited by our time on field (see Table 1) and by the criteria and means we use to identify flares (Section 3).

This sample of short duration flares contains sources across a broad range of flare amplitudes ($\Delta M$ 0.1 − 1.8 mag), as well as across the full range of distances out to 500 pc, see panel (b) and (c) of Figure 13. In Figure 13 (b), there are visible vertical lines along which flare durations cluster, an artifact of our continuous 20 second observations which cause the durations to artificially be distributed in multiples of ∼50 second intervals (20 s exposures and ∼30 s readout). It is also important to note that the true flare duration values were not extrapolated for the cut off flares in our sample (those with minimum duration times as a result of the limited time on field).

Short duration flares have previously been identified by (Brasseur et al. 2019) within the Galaxy Evolution Explorer (GALEX Bianchi & GALEX Team 1999) mission's 10-second cadence data. The bulk of the stars observed by GALEX were within the mid-F to mid-K spectral type range. Our work confirms that this short duration population does indeed continue into the late-K to mid M-dwarf stellar types, shown in Figure 13 (c).

Our sample of short duration flares spans a considerable energy range of $10^{31}$–$10^{34}$ ergs, further diversifying the overall population of short duration flares beyond the GALEX sample. Interestingly, lower energy flares ($10^{33}$ erg) are not limited to only short duration events, see Figure 7.

### 4.6 Flare Frequency Distribution

The Flare Frequency Distribution (FFD) represents the occurrence frequency ($dN/dE$) of flares as a function of flare energy (E), and can be expressed as a power-law relation i.e. $dN/dE \sim E^{-\alpha}$. Previous work, including that on Solar flares and Kepler flares, find generally $\alpha$ is ∼2, with slight variations in the value of $\alpha$ found across individual spectral types (Shibata et al. 2013; Maehara et al. 2012; Shibayama et al. 2013; Yang et al. 2017; Yang & Liu 2019). For example, Audard et al. (2000) found that the FFD became flat towards later spectral types, however Yang & Liu (2019) findings indicate an increase in FFD across M dwarfs. Davenport et al. (2019) noted that comparing stellar types using specific flare rates generated from an individual FFD implicitly assumes that the flare rate for a star is governed by a single power law at all energies examined. This can be useful in the case of comparing samples with different observing conditions, however Davenport et al. (2019) further commented that in doing so we must ensure that a sufficient number of flares are observed in each spectral range to adequately measure the power law distribution within the FFD. With this in mind, we have not separated our sample to calculate FFDs for each individual spectral types, rather calculating $\alpha$ from an FFD of the combined sample to ensure a sufficient number of flares in each energy bin. The FFD of the 96 flares in our sample, shown in Figure 14, is fit with the least-squares power law of $log(dN/dE) = E^\alpha$. We find a value of $\alpha \sim 1.43 \pm 0.37$, in agreement with previous $\alpha$ estimates across both solar-like and dwarf stars (e.g. Audard et al. 2000; Maehara et al. 2015; Yang & Liu 2019), suggesting that the large portion of short duration flares found in this work are likely produced by the same physical dynamo mechanism. Our FFD does however suggests that higher energy flares occur less frequently amongst this paper's sample in comparison to previous work. Once the entirety of the archival DWF optical data is mined for flares, producing a full magnitude limited catalog, FFDs for each spectral type can be produced. This future work will assist with current research into understanding flare mechanisms between magnetically weak giants and magnetically active dwarf stars.

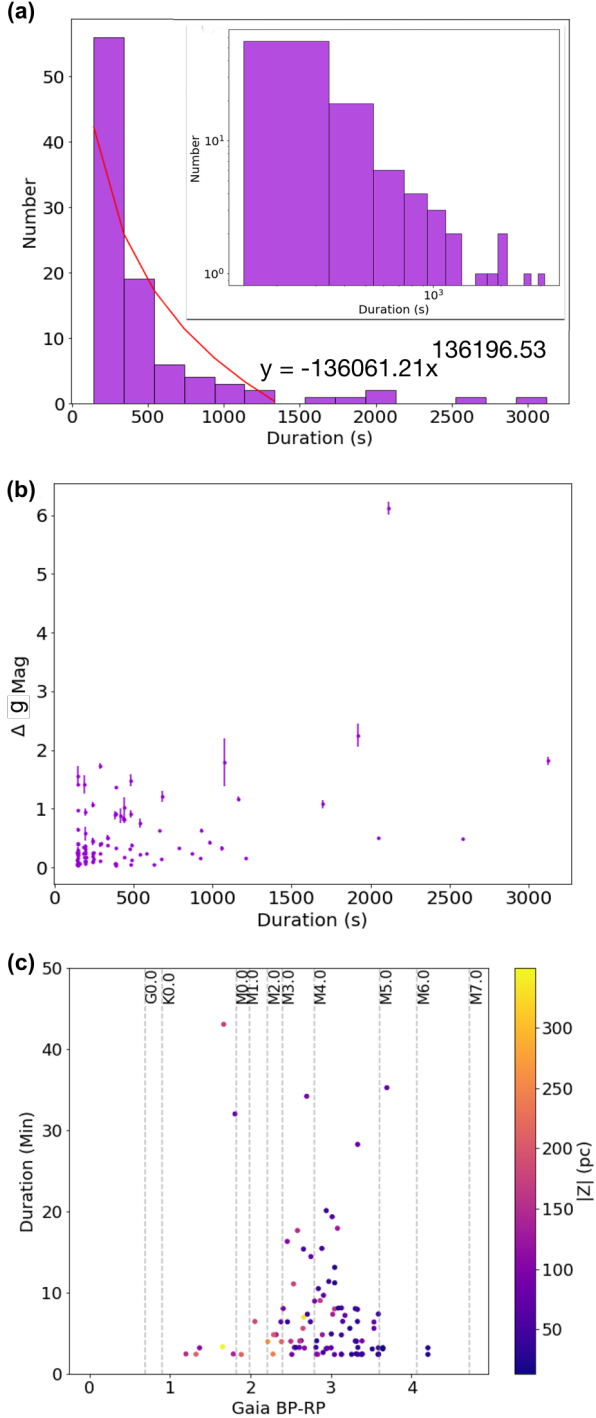

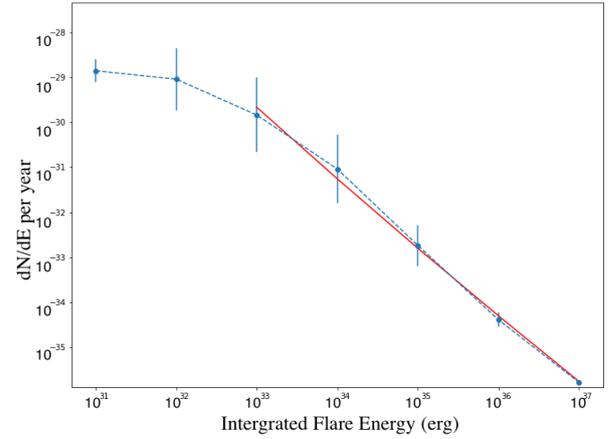

**Figure 14.** The Flare Frequency Distribution (FFD) for all flares discovered in this work. The FFD spans the energy range of $10^{31}$ to $10^{37}$. The error bar of each energy bin is calculated assuming Poisson statistics. The red line indicates the least-squares power law of $\log(dN/dE)$, finding $\alpha \sim 1.43 \pm 0.37$, fitted using data between $10^{33}$ to $10^{37}$.

### 4.7  Flare Rates

Across our 12 target fields and within 500 pc, we find that the number of flares per hour varies considerably, having an average of $0.5 \pm 0.2$ flares $deg^{-2}$ $hr^{-1}$, see Table 3. Our daily flare rate of 10.54 flares $deg^{-2}$ $day^{-1}$ is also comparable to the daily rate of 7.6 flares $deg^{-2}$ $day^{-1}$ on sources of 3500K $\leq T_{eff} \leq$ 4000K across all flare energies from (Feinstein et al. 2020).

We further breakdown flare rates across the range of integrated flare energies, as shown in Table 4. We find the highest rates are associated with the lowest energy flares as expected and in agreement with previous studies (Maehara et al. 2012; Shibayama et al. 2013; Yang et al. 2017; Yang & Liu 2019; Günther et al. 2019). Using the combined rates across all observed flare energies in this study, we estimate a sky rate of ∼0.43 events deg$^{-2}$ hr$^{-1}$ and volumetric rate of ∼2.9±0.3 × 10$^{-6}$ flares pc$^3$ hr$^{-1}$, see Tables 3 and 5 respectively. Note: our rates are likely an underestimate of the true flare rate, a reflection of selection criteria and observational limitations.

**Figure 13.** *Top* a) Histogram of flare duration. The majority of flares occur over short durations, following an exponential relationship. *Middle* b) Change in magnitude (flare amplitude) versus duration, with short duration events spanning from 0.1–1.8 magnitudes. The majority of flares have < 2 'g' band magnitude change, with one notable exception present at ∼ 6 magnitudes. *Bottom* c) Duration (minutes) versus quiescent source colour (Gaia BP-RP). The colour of each point represents |Z| (pc) as indicated by the colour bar. Short duration events are present across multiple spectral types out to 500 pc.

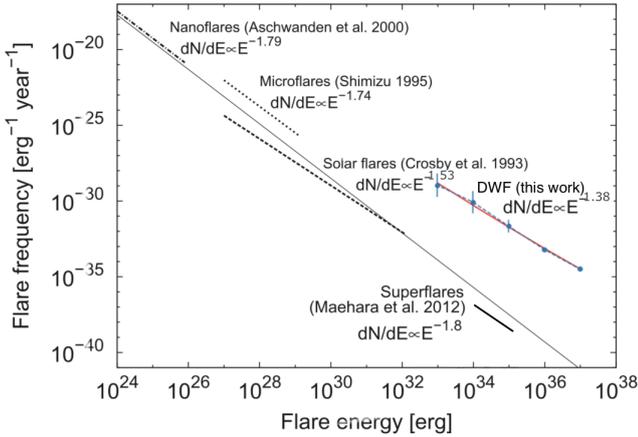

**Figure 15.** The Flare Frequency Distribution (FFD) for the DWF sample of flares compared to several previous studies on solar type flares.

## 5  CONCLUSIONS

We present our 500 pc distance-limited stellar flare study, using DECam continuous 20 second cadence *g* band data collected over 12 fields via the Deeper, Wider, Faster program (DWF). The advantage of our choice of continuous blue filter observations is our sensitivity to low-amplitude flares, where traditionally the detection efficiency drops at redder wavelengths. Furthermore, the unique observation strategy of DWF provides a 20-second cadence sensitive to uncovering fast evolving short duration flares. We search the light curves of 19,914 sources, each with multiple nightly light curves. The average time per visit was ∼74 minutes on each 2.2° diameter DECam field.

We find 96 flare events, occurring across 80 stars, from our volume limited sample of 19,914 sources. We found the following:

• Flares were found across K– M6 type stars, strongly concentrated around the mid type M-dwarfs.

• The flares identified vary greatly in both duration and flare energy. Duration's span from ∼2.4 to ∼52 minutes, with ∼ 70% having duration's < 8 minutes.

• The short duration flares identified in this work span vast energy distribution of $10^{31} - 10^{37}$ erg.

• We confirm a strong relationship between flare energy and distance from the Galactic plane, with lower energy flares occurring ∼<100 pc. This is consistent with the estimated scale height of the Galactic young disk.

• The flaring fraction of stars is also identified to decrease with vertical distance from the Galactic plane, in agreement with previous studies.

• We determine, using our sources transverse velocity, that our flares are generated from likely young stellar sources.

• Our FFD confirms a power law relationship between flare number and flare energy, and generates a value of $\alpha$ is consistent with previous work.

• We determine an average flare rate of 0.42 ± 0.2 flares deg$^{-2}$ hr$^{-1}$, and a volumetic rate of 2.9 ± 0.3 × 10$^{-6}$ flares pc$^{-3}$ hr$^{-1}$.

We plan to continue to explore flare activity within the DWF optical data sets, aiming to provide a full magnitude limited flare catalog. DWF will also aim to explore multi-wavelength properties of flares via past and future coordinated DWF simultaneous observations.

## ACKNOWLEDGEMENTS

We'd like to acknowledge and thank our reviewer for their very insight and helpful review and comments. Part of this research was funded by the Australian Research Council Centre of Excellence for Gravitational Wave Discovery (Oz-Grav), CE170100004. JC acknowledges funding from the Australian Research Council Discovery Project, DP200102102. We acknowledge the financial assistance of the National Research Foundation (NRF). Opinions expressed and conclusions arrived at, are those of the authors and are not necessarily to be attributed to the NRF. This work was partly supported by the GROWTH (Global Relay of Observatories Watching Transients Happen) project funded by the National Science Foundation under PIRE Grant No 1545949. S.A.B. acknowledges support from the Aliyun Fellowship and Chinese Academy of Sciences President's International Fellowship Initiative Grant. This work has made use of data from the European Space Agency (ESA) mission Gaia (https://www.cosmos.esa. int/gaia), processed by the Gaia Data Processing and Analysis Consortium (DPAC, https://www.cosmos.esa. int/web/gaia/dpac/consortium). Funding for the DPAC has been provided by national institutions, in particular the institutions participating in the Gaia Multilateral Agreement.This publication makes use of data products from the Wide-field Infrared Survey Explorer, which is a joint project of the University of California, Los Angeles, and the Jet Propulsion Laboratory/California Institute of Technology, funded by the National Aeronautics and Space Administration. This project used data obtained with the Dark Energy Camera (DECam), which was constructed by the Dark Energy Survey (DES) collaboration. Funding for the DES Projects has been provided by the U.S. Department of Energy, the U.S. National Science Foundation, the Ministry of Science and Education of Spain, the Science and Technology Facilities Council of the United Kingdom, the Higher Education Funding Council for England, the National Centre for Supercomputing Applications at the University of Illinois at Urbana-Champaign, the Kavli Institute of Cosmological Physics at the University of Chicago, the Center for Cosmology and Astro-Particle Physics at the Ohio State University, the Mitchell Institute for Fundamental Physics and Astronomy at Texas A&M University, Financiadora de Estudos e Projetos, Fundação Car-

| Field | Flares (#) | Flare Sources (#) | Time on Field (hrs) | # Flares (deg$^{-2}$ hr$^{-1}$) | Galactic Latitude (deg) |
|---|---|---|---|---|---|
| Antlia | 8 | 8 | 7.03 | 0.39 | 19.172 |
| Dusty10 | 4 | 4 | 3.50 | 0.39 | −19.957 |
| Dusty12 | 3 | 3 | 4.39 | 0.23 | −21.889 |
| FRB131104 | 13 | 11 | 6.37 | 0.70 | −21.93 |
| 8hr | 6 | 6 | 4.95 | 0.41 | −22.618 |
| Dusty11 | 1 | 1 | 2.35 | 0.14 | −22.814 |
| NGC6744 | 27 | 18 | 18.38 | 0.50 | −26.054 |
| Prime | 16 | 16 | 6.88 | 0.80 | −30.262 |
| NSF2 | 3 | 3 | 3.61 | 0.28 | −39.823 |
| FRB010724 | 6 | 5 | 2.63 | 0.78 | −41.804 |
| 4hr | 10 | 7 | 8.05 | 0.43 | −44.756 |
| 3hr | 9 | 7 | 6.88 | 0.44 | −53.432 |

**Table 3.** The number of flares and flaring sources recorded across all fields, with average hourly rates calculated based on the DECam 2.91 squared degree field of view.

| Int. Flare Energy (ergs) | # Flares | Rate (Deg$^{-2}$ Hr$^{-1}$) | Rate (Deg$^{-2}$ Day$^{-1}$) | Rate (Deg$^{-2}$ Yr$^{-1}$) | Rate (Sky$^{-1}$ Yr$^{-1}$) |
|---|---|---|---|---|---|
| $10^{31}$ | 3 | 0.014 | 0.32 | 120.3 | 4.96 ×10$^6$ |
| $10^{32}$ | 20 | 0.091 | 2.19 | 801.9 | 3.31 ×10$^7$ |
| $10^{33}$ | 36 | 0.164 | 3.95 | 1443.4 | 5.95 ×10$^7$ |
| $10^{34}$ | 27 | 0.123 | 2.96 | 10082.6 | 4.47 ×10$^7$ |
| $10^{35}$ | 7 | 0.032 | 0.77 | 280.7 | 1.16 ×10$^6$ |
| $10^{36}$ | 2 | 0.009 | 0.21 | 80.2 | 3.31 ×10$^6$ |
| $10^{37}$ | 1 | 0.004 | 0.11 | 40.1 | 1.65 ×10$^6$ |
| Total | 96 | 0.43 | 10.54 | 3849 | 1.59 ×10$^8$ |

**Table 4.** Flare rates determined for individual energy ranges with hourly and yearly rates calculate based on the DECam 2.91 squared degree field of view.

| Δ g Magnitude of Flare | # Flares | # Flares pc$^3$ hr$^{-1}$ |
|---|---|---|
| All | 96 | 2.88×10$^{-6}$ |
| (0.0,0.5] | 65 | 1.90×10$^{-6}$ |
| (0.5, 1.0] | 16 | 4.81×10$^{-7}$ |
| (1.0,1.5] | 9 | 2.70×10$^{-7}$ |
| (1.5 ,2.0] | 4 | 1.20×10$^{-7}$ |
| >2.0 | 2 | 6.01×10$^{-8}$ |

**Table 5.** Volumetic flare rates pc$^3$ per hour. Calculated using 75.08 hours of observations and a total observational volume of 443218 pc$^3$ for the 12 fields out to 500 pc.

## DATA AVAILABILITY

The data underlying this article will be shared on reasonable request to the corresponding author.

## REFERENCES

Abazajian K. N., et al., 2009, ApJS, 182, 543
Andreoni I., Cooke J., 2018, arXiv e-prints, p. arXiv:1802.01100
Andreoni I., et al., 2020, MNRAS, 491, 5852
Audard M., Güdel M., Drake J. J., Kashyap V. L., 2000, ApJ, 541, 396
Berger E., et al., 2008, ApJ, 676
Bertin E., Arnouts S., 1996, A&AS, 117, 393
Bertin E., Arnouts S., 2010, SExtractor: Source Extractor (ascl:1010.064)
Bianchi L., GALEX Team 1999, Mem. Soc. Astron. Italiana, 70, 365
Binney J., Tremaine S., 1987, Galactic dynamics
Bopp B. W., Moffett T. J., 1973, ApJ, 185, 239
Borucki W. J., et al., 2010, Science, 327, 977
Brasseur C. E., Osten R. A., Fleming S. W., 2019, ApJ, 883, 88
Cabrera-Lavers A., Bilir S., Ak S., Yaz E., López-Corredoira M., 2007, A&A, 464, 565
Casagrande L., et al., 2020, arXiv e-prints, p. arXiv:2011.02517
Chabrier G., Baraffe I., 1997, A&A, 327, 1039
Chang S. W., Byun Y. I., Hartman J. D., 2015, ApJ, 814, 35
Chang S.-W., Wolf C., Onken C. A., 2020, MNRAS, 491, 39
Cutri R. M., et al., 2021, VizieR Online Data Catalog, p. II/328
Dal H. A., Evren S., 2010, AJ, 140, 483
Davenport J. R. A., 2016, in Kosovichev A. G., Hawley S. L., Heinzel P., eds, IAU Symposium Vol. 320, Solar and Stellar Flares and their Effects on Planets. pp 128–133 (arXiv:1510.05695), doi:10.1017/S174392131600867X
Davenport J. R. A., Becker A. C., Kowalski A. F., Hawley S. L., Schmidt S. J., Hilton E. J., Sesar B., Cutri R., 2012, ApJ, 748, 58
Davenport J. R. A., Covey K. R., Clarke R. W., Boeck A. C., Cornet J., Hawley S. L., 2019, ApJ, 871, 241
Dillon C. J., et al., 2020, ApJ, 904, 109
Dorman B., Nelson L. A., Chau W. Y., 1989, ApJ, 342, 1003
Ducati J. R., Bevilacqua C. M., Rembold S. B., Ribeiro D., 2001, ApJ, 558, 309
Evans D. W., et al., 2018, A&A, 616, A4
Feinstein A. D., Montet B. T., Ansdell M., Nord B., Bean J. L., Günther M. N., Gully-Santiago M. A., Schlieder J. E., 2020, AJ, 160, 219
Flaugher B., et al., 2015, The Astronomical Journal, 150, 150
Furusawa H., et al., 2018, PASJ, 70, S3
Garnavich P. M., et al., 2016, ApJ, 820
Günther M. N., et al., 2019, arXiv e-prints, p. arXiv:1901.00443
Haisch B., Strong K. T., Rodono M., 1991, Annual Review of Astronomy and Astrophysics, 29, 275
Hawley S. L., Pettersen B. R., 1991, ApJ, 378, 725
Hawley S. L., Davenport J. R. A., Kowalski A. F., Wisniewski J. P., Hebb L., Deitrick R., Hilton E. J., 2014, ApJ, 797, 121
Hilton E. J., et al., 2010, AJ, 140, 1402
Howard W. S., et al., 2018, ApJ, 860, L30
Howard W. S., et al., 2020, ApJ, 902, 115
Ishida K., Ichimura K., Shimizu Y., Mahasenaputra 1991, Ap&SS, 182, 227
Jackman J. A. G., et al., 2018, MNRAS, 477, 4655
Jurić M., et al., 2008, ApJ, 673, 864
Kawanomoto S., et al., 2018, PASJ, 70, 66
Komiyama Y., et al., 2018, PASJ, 70, S2
Kowalski A. F., et al., 2009, AJ, 138, 633

Lacy C. H., Moffett T. J., Evans D. S., 1976, The Astrophysical Journal Supplement Series, 30, 85
Law N. M., Fors O., Wulfken P., Ratzloff J., Kavanaugh D., 2014, in Stepp L. M., Gilmozzi R., Hall H. J., eds, Society of Photo-Optical Instrumentation Engineers (SPIE) Conference Series Vol. 9145, Ground-based and Airborne Telescopes V. p. 91450Z (arXiv:1407.0026), doi:10.1117/12.2057031
Lin R. P., Hudson H. S., 1976, Sol Physics, 50, 153
Lochner M., Bassett B. A., 2020, arXiv e-prints, p. arXiv:2010.11202
Luri X., et al., 2018, A&A, 616, A9
Mackereth J. T., et al., 2019, MNRAS, 489, 176
Maehara H., et al., 2012, Nature, 485, 478
Maehara H., Shibayama T., Notsu Y., Notsu S., Honda S., Nogami D., Shibata K., 2015, Earth, Planets, and Space, 67, 59
Miyazaki S., et al., 2018, PASJ, 70, S1
Moffett T. J., 1974, The Astrophysical Journal Supplement Series, 29, 1
Mondrik N., Newton E., Charbonneau D., Irwin J., 2019, ApJ, 870, 10
Morales J. C., et al., 2009, ApJ, 691, 1400
Mould J., Clementini G., Da Costa G., 2019, Publ. Astron. Soc. Australia, 36, e001
Namekata K., et al., 2017, ApJ, 851, 91
Onken C. A., et al., 2019, Publ. Astron. Soc. Australia, 36, e033
Osten R. A., Wolk S. J., 2015, The Astrophysical Journal, 809, 79
Osten R. A., et al., 2005, ApJ, 621, 398
Parnell C. E., Jupp P. E., 2000, ApJ, 529, 554
Perley D. A., et al., 2018, Monthly Notices of the Royal Astronomical Society, 484, 1031
Pettersen B. R., 1989, Sol Physics, 121, 299
Pineda J. S., et al., 2013, Astronomical Journal, 146
Prentice S. J., et al., 2018, ApJ, 865, L3
Ricker G. R., et al., 2009, in American Astronomical Society Meeting Abstracts #213. p. 403.01
Rix H.-W., Bovy J., 2013, The Astronomy and Astrophysics Review, 21, 61
Robin A. C., et al., 2012, A&A, 543, A100
Rodríguez Martínez R., Lopez L. A., Shappee B. J., Schmidt S. J., Jayasinghe T., Kochanek C. S., Auchettl K., Holoien T. W. S., 2020a, ApJ, 892, 144
Rodríguez Martínez R., Lopez L. A., Shappee B. J., Schmidt S. J., Jayasinghe T., Kochanek C. S., Auchettl K., Holoien T. W. S., 2020b, ApJ, 892, 144
Schmidt S. J., Wagoner E. L., Johnson J. A., Davenport J. R. A., Stassun K. G., Souto D., Ge J., 2016, Monthly Notices of the Royal Astronomical Society, 460, 2611
Schmidt S. J., et al., 2018, arXiv e-prints, p. arXiv:1809.04510
Scott D., Pierfederici F., Swaters R. A., Thomas B., Valdes F. G., 2007, in Shaw R. A., Hill F., Bell D. J., eds, Astronomical Society of the Pacific Conference Series Vol. 376, Astronomical Data Analysis Software and Systems XVI. p. 265
Seabroke G. M., Gilmore G., 2007, Monthly Notices of the Royal Astronomical Society, 380, 1348
Shappee B., et al., 2012, in American Astronomical Society Meeting Abstracts #220. p. 432.03
Shibata K., et al., 2013, PASJ, 65, 49
Shibayama T., et al., 2013, The Astrophysical Journal Supplement Series, 209, 5
Skumanich A., 1972, ApJ, 171, 565
Stelzer B., Damasso M., Scholz A., Matt S. P., 2016, MNRAS, 463, 1844
Swaters R. A., Valdes F. G., 2007, in Shaw R. A., Hill F., Bell D. J., eds, Astronomical Society of the Pacific Conference Series Vol. 376, Astronomical Data Analysis Software and Systems XVI. p. 269

Ting  K.,  Liu  F.,  Zhou  Z.,  2008,  in  ICDM  2008.
    Eighth  IEEE  International  Conference  on  Data  Min-
    ing.   IEEE   Computer   Society,   Los   Alamitos,   CA,
    USA,   pp   413–422,   doi:10.1109/ICDM.2008.17,   https:
    //doi.ieeecomputersociety.org/10.1109/ICDM.2008.17

Trumpler R. J., Weaver H. F., 1953, Statistical astronomy [by]
    Robert J. Trumpler and Harold F. Weaver. University of Cal-
    ifornia Press Berkeley

Valdes F. G., Swaters R. A., 2007, in Shaw R. A., Hill F., Bell
    D. J., eds, Astronomical Society of the Pacific Conference Se-
    ries Vol. 376, Astronomical Data Analysis Software and Sys-
    tems XVI. p. 273

Van Doorsselaere T., Shariati H., Debosscher J., 2017, ApJS, 232,
    26

Walkowicz L. M., et al., 2011, AJ, 141, 50

Webb S., et al., 2020, MNRAS, 498, 3077

West A. A., Hawley S. L., 2008, PASP, 120, 1161

West A. A., et al., 2004, AJ, 128, 426

West A. A., Hawley S. L., Bochanski J. J., Covey K. R., Reid
    I. N., Dhital S., Hilton E. J., Masuda M., 2008, AJ, 135, 785

Wheatley P. J., et al., 2018, MNRAS, 475, 4476

Wolf C., et al., 2018, Publ. Astron. Soc. Australia, 35, e010

Wright N. J., Drake J. J., Mamajek E. E., Henry G. W., 2011,
    The Astrophysical Journal, 743, 48

Yang H., Liu J., 2019, ApJS, 241, 29

Yang H., et al., 2017, ApJ, 849, 36

**APPENDIX A: DWF 500 PC FLARES**

| Date | Gaia DR2 ID | Quies-cent Mag (g) | Gaia BP-RP | WISE W1-W2 | SM DR2/3 r-z | Flare Δ Mag (g) | Flare Dura-tion $^a$ (s) | Dist-ance (pc) | Quies-cent Flux (erg/sec) | Rela-tive Flux ($\Sigma$) | Flare Energy $^b$ (g-band) (erg) | Flare Energy $^b$ (bol) (erg) |
|---|---|---|---|---|---|---|---|---|---|---|---|---|
| 151218 | 4636450838513006208 | 17.63 | 2.95 | 0.19 | 2.57 | 0.17 | 189.48 | 73.80 | $1.25\times10^{30}$ | 0.32 | $7.51\times10^{31}$ | $6.91\times10^{32}$ |
| 151222 | 4636433108887748864 | 17.26 | 3.23 | 0.19 | 2.75 | 0.38 | 336.35 | 39.90 | $5.39\times10^{29}$ | 0.80 | $1.45\times10^{32}$ | $1.33\times10^{33}$ |
| 151218 | 4636360850358017792 | 17.61 | 2.94 | 0.20 | 2.48 | 0.16 | 1206.85+ | 61.94 | $9.48\times10^{29}$ | 1.41 | $1.61\times10^{33}+$ | $1.48\times10^{34}$ |
| 151219 | 4637865394582180480 | 16.44 | 3.33 | 0.17 | 2.90 | 1.36 | 476.49 | 50.75 | $1.85\times10^{30}$ | 0.80 | $6.48\times10^{32}$ | $5.96\times10^{33}$ |
| 151220 | 4637865394582180480 | 16.44 | 3.33 | 0.17 | 2.90 | 1.36 | 386.81 | 50.75 | $1.85\times10^{30}$ | 4.04 | $2.89\times10^{33}$ | $2.66\times10^{34}$ |
| 151218 | 4728699589203817600 | 18.37 | 3.13 | 0.21 | 2.52 | 0.25 | 144.22 | 58.26 | $3.85\times10^{29}$ | 0.22 | $1.23\times10^{31}$ | $1.13\times10^{32}$ |
| 151218 | 4728699589203817600 | 18.37 | 3.13 | 0.21 | 2.52 | 0.24 | 287.72+ | 58.26 | $3.85\times10^{29}$ | 0.53 | $5.87\times10^{31}+$ | $5.40\times10^{32}$ |
| 151221 | 4728699589203817600 | 18.39 | 3.13 | 0.21 | 2.52 | 0.64 | 144.41 | 58.26 | $3.82\times10^{29}$ | 1.06 | $5.89\times10^{31}$ | $5.42\times10^{32}$ |
| 151222 | 4728699589203817600 | 18.36 | 3.13 | 0.21 | 2.52 | 0.38 | 487.09 | 58.26 | $4.17\times10^{29}$ | 1.34 | $2.57\times10^{32}$ | $2.37\times10^{33}$ |
| 151218 | 4728703055241994752 | 20.94 | 3.69 | 0.17 | 3.61 | 6.12 | 2115.37+ | 97.35 | $1.08\times10^{29}$ | 1105.8 | $2.38\times10^{35}+$ | $2.19\times10^{36}$ |
| 151219 | 4734505525995537024 | 18.03 | 3.32 | 0.18 | 2.80 | 0.09 | 240.32 | 47.31 | $3.73\times10^{29}$ | 0.15 | $1.35\times10^{31}$ | $1.24\times10^{32}$ |
| 151220 | 4683390089413218688 | 19.93 | 4.20 | 0.23 | 3.32 | 0.97 | 143.69 | 38.54 | $2.74\times10^{29}$ | 0.33 | $2.02\times10^{32}$ | $1.85\times10^{33}$ |
| 151220 | 4683390089413218688 | 19.93 | 4.20 | 0.23 | 3.32 | 0.94 | 192.29 | 38.54 | $2.74\times10^{29}$ | 2.01 | $1.65\times10^{31}$ | $1.52\times10^{32}$ |
| 150114 | 4779473318188791680 | 16.84 | 3.05 | 0.20 | 2.51 | 0.14 | 672.56 | 38.51 | $7.38\times10^{29}$ | 0.33 | $1.64\times10^{32}$ | $1.51\times10^{33}$ |
| 150117 | 4779473318188791680 | 16.84 | 3.05 | 0.20 | 2.51 | 0.33 | 786.71 | 38.51 | $7.38\times10^{29}$ | 0.91 | $4.95\times10^{32}$ | $4.55\times10^{33}$ |
| 151221 | 4779669134337892864 | 18.85 | 3.05 | 0.19 | 2.49 | 0.13 | 142.84 | 97.66 | $7.52\times10^{29}$ | 0.21 | $2.06\times10^{31}$ | $1.90\times10^{32}$ |
| 170205 | 5502472749699352320 | 17.17 | 2.82 | 0.18 | 2.28 | 0.14 | 243.49 | 75.21 | $3.07\times10^{30}$ | 0.31 | $1.56\times10^{32}$ | $1.44\times10^{33}$ |
| 150114 | 5502105963787924480 | 17.98 | 3.64 | 0.22 | 3.16 | 0.09 | 193.21 | 34.35 | $1.94\times10^{29}$ | 0.19 | $7.11\times10^{30}$ | $6.54\times10^{31}$ |
| 150117 | 5502105963787924480 | 17.98 | 3.64 | 0.22 | 3.16 | 0.35 | 184.04 | 34.35 | $1.94\times10^{29}$ | 0.56 | $2.01\times10^{31}$ | $1.85\times10^{32}$ |
| 170203 | 5447196146939389568 | 19.31 | 3.09 | 0.19 | 2.57 | 0.12 | 485.75 | 81.13 | $3.37\times10^{29}$ | 0.60 | $9.84\times10^{31}$ | $9.05\times10^{32}$ |
| 170206 | 5444145826805790080 | 19.12 | 3.31 | 0.21 | 2.70 | 0.10 | 147.62 | 98.85 | $5.60\times10^{29}$ | 3.16 | $2.62\times10^{32}$ | $2.41\times10^{33}$ |
| 160802 | 6435079169511271040 | 19.78 | 3.30 | 0.18 | 2.89 | 0.91 | 482.12 | 97.22 | $2.89\times10^{29}$ | 0.16 | $2.30\times10^{31}$ | $2.11\times10^{32}$ |
| 160803 | 6435079169511271040 | 19.92 | 3.30 | 0.18 | 2.89 | 0.47 | 242.23 | 97.22 | $2.89\times10^{29}$ | 0.71 | $7.08\times10^{31}$ | $6.51\times10^{32}$ |
| 160803 | 6438591692548589056 | 18.91 | 3.33 | 0.25 | 2.80 | 0.07 | 194.56 | 85.74 | $1.18\times10^{29}$ | 0.20 | $2.14\times10^{31}$ | $1.97\times10^{32}$ |
| 160804 | 6438591692548589056 | 18.91 | 3.33 | 0.25 | 2.80 | 0.21 | 144.22 | 85.74 | $1.18\times10^{29}$ | 0.39 | $1.25\times10^{32}$ | $1.15\times10^{33}$ |
| 160728 | 6435402765228918144 | 20.00 | 3.59 | 0.23 | 3.12 | 0.22 | 143.77 | 80.42 | $1.72\times10^{29}$ | 0.18 | $2.68\times10^{31}$ | $2.47\times10^{32}$ |
| 160802 | 6435402765228918144 | 20.05 | 3.59 | 0.23 | 3.12 | 0.36 | 192.16 | 80.41 | $1.72\times10^{29}$ | 0.58 | $1.94\times10^{31}$ | $1.78\times10^{32}$ |
| 160802 | 6435402765228918144 | 20.05 | 3.59 | 0.23 | 3.12 | 0.22 | 144.37 | 80.41 | $1.72\times10^{29}$ | 0.41 | $1.01\times10^{31}$ | $9.31\times10^{31}$ |
| 160803 | 6435402765228918144 | 20.05 | 3.59 | 0.23 | 3.12 | 0.30 | 144.63 | 80.41 | $1.72\times10^{29}$ | 0.51 | $1.05\times10^{31}$ | $1.39\times10^{33}$ |
| 160807 | 6435402765228918144 | 20.05 | 3.59 | 0.23 | 3.12 | 0.81 | 442.08 | 80.41 | $1.72\times10^{29}$ | 3.35 | $2.55\times10^{32}$ | $2.34\times10^{33}$ |
| 160804 | 6435402700808342528 | 14.71 | 2.46 | 0.11 | 1.87 | 0.04 | 384.66 | 79.89 | $2.26\times10^{31}$ | 0.22 | $1.94\times10^{33}$ | $1.78\times10^{33}$ |
| 160805 | 6435404453155002624 | 18.94 | 3.38 | 0.21 | 3.00 | 0.41 | 144.12 | 94.09 | $6.37\times10^{29}$ | 0.90 | $8.27\times10^{31}$ | $7.60\times10^{32}$ |
| 160805 | 6435404453155002624 | 18.94 | 3.38 | 0.21 | 3.00 | 1.41 | 143.94 | 94.09 | $6.37\times10^{29}$ | 3.59 | $3.29\times10^{32}$ | $3.03\times10^{33}$ |
| 160806 | 6438625781704284928 | 16.75 | 2.84 | 0.18 | 2.33 | 0.06 | 629.72 | 94.79 | $4.90\times10^{30}$ | 0.41 | $1.26\times10^{33}$ | $1.16\times10^{34}$ |
| 150117 | 4758284595250930560 | 19.09 | 3.51 | 0.19 | 2.95 | 0.25 | 183.69 | 60.58 | $2.15\times10^{29}$ | 0.44 | $1.76\times10^{31}$ | $1.61\times10^{32}$ |
| 170202 | 4758503569863047424 | 16.03 | 2.66 | 0.17 | 2.09 | 0.17 | 922.50+ | 96.63 | $9.22\times10^{30}$ | 0.96 | $8.17\times10^{33}+$ | $7.52\times10^{34}$ |
| 170202 | 4758620770931164544 | 17.63 | 2.90 | 0.16 | 2.37 | 0.07 | 157.89 | 89.61 | $1.81\times10^{30}$ | 0.10 | $2.97\times10^{31}$ | $2.74\times10^{32}$ |
| 160730 | 5191257877539468672 | 17.41 | 3.25 | 0.23 | 2.75 | 0.08 | 144.29 | 69.57 | $1.42\times10^{30}$ | 0.11 | $2.15\times10^{31}$ | $1.98\times10^{32}$ |
| 151218 | 4728688628448657280 | 17.69 | 2.51 | 0.17 | 1.88 | 0.25 | 143.49 | 113.05 | $2.74\times10^{30}$ | 0.29 | $1.17\times10^{32}$ | $1.07\times10^{33}$ |
| 151222 | 4728701135393440384 | 16.62 | 1.65 | -0.04 | 0.82 | 0.18 | 200 | 435.55 | $1.09\times10^{32}$ | 0.21 | $4.55\times10^{33}$ | $4.19\times10^{34}$ |
| 151219 | 4683671426951508864 | 18.61 | 3.53 | 0.21 | 3.10 | 0.51 | 336.54 | 108.14 | $1.07\times10^{30}$ | 1.85 | $6.67\times10^{32}$ | $4.38\times10^{33}$ |
| 150116 | 4683671426951508864 | 18.61 | 3.53 | 0.21 | 3.10 | 0.89 | 383.17 | 108.14 | $1.07\times10^{30}$ | 3.04 | $4.76\times10^{32}$ | $6.14\times10^{33}$ |
| 150114 | 4683434958935655552 | 22.19 | 2.66 | 0.26 | 2.40 | 0.88 | 418.77 | 476.23 | $7.70\times10^{29}$ | 2.78 | $8.98\times10^{32}$ | $8.26\times10^{33}$ |
| 150115 | 4779763005142093696 | 18.62 | 2.21 | 0.14 | 1.35 | 0.1 | 237.83 | 384.70 | $1.35\times10^{31}$ | 0.28 | $9.24\times10^{32}$ | $8.50\times10^{33}$ |
| 150116 | 4778723996718806784 | 20.57 | 2.86 | 0.18 | 2.44 | 0.75 | 541.34 | 233.38 | $8.23\times10^{29}$ | 1.53 | $6.81\times10^{32}$ | $6.27\times10^{33}$ |
| 180606 | 5209798289282962048 | 15.71 | 1.66 | 0.03 | 0.94 | 0.49 | 2584.01 | 453.82 | $2.73\times10^{32}$ | 13.27 | $9.37\times10^{36}$ | $8.62\times10^{37}$ |
| 180606 | 5209089821541031680 | 18.74 | 2.54 | 0.16 | 1.93 | 0.27 | 195.22 | 287.84 | $6.75\times10^{30}$ | 0.34 | $4.52\times10^{32}$ | $4.16\times10^{33}$ |
| 180609 | 5209799457513998720 | 17.53 | 2.81 | 0.19 | 2.29 | 0.23 | 195.16 | 101.59 | $2.56\times10^{30}$ | 0.58 | $2.92\times10^{32}$ | $2.69\times10^{33}$ |
| 180609 | 5209096246811787904 | 22.06 | 3.08 | 0.13 | 2.42 | 1.79 | 1075.85 | 328.14 | $4.12\times10^{29}$ | 10.20 | $4.52\times10^{33}$ | $4.16\times10^{34}$ |
| 180606 | 5209076425539832448 | 19.59 | - | 0.16 | 2.25 | 1.82 | 3121.58 | 404.49 | $6.09\times10^{30}$ | 51.91 | $9.87\times10^{35}$ | $9.08\times10^{36}$ |
| 170203 | 5447129489047169536 | 19.14 | 2.79 | 0.13 | 2.36 | 0.22 | 537.92 | 263.78 | $3.92\times10^{30}$ | 1.41 | $2.98\times10^{33}$ | $2.74\times10^{34}$ |
| 180608 | 5444073018520551296 | 19.46 | 3.14 | 0.23 | 2.67 | 0.91 | 387.53 | 173.67 | $1.27\times10^{30}$ | 1.89 | $9.28\times10^{32}$ | $8.54\times10^{33}$ |
| 170203 | 5444224407527438208 | 17.21 | 1.79 | -0.01 | 0.93 | 0.04 | 146.33 | 423.10 | $5.97\times10^{30}$ | 0.07 | $6.45\times10^{32}$ | $5.93\times10^{33}$ |
| 170207 | 5445901884314615680 | 18.87 | 2.46 | 0.11 | 1.86 | 0.43 | 978.76 | 323.55 | $7.57\times10^{30}$ | 2.89 | $2.14\times10^{34}$ | $1.97\times10^{35}$ |
| 170205 | 5444220314420785280 | 20.13 | 2.60 | 0.21 | 2.09 | 0.58 | 194.93 | 232.57 | $1.22\times10^{30}$ | 1.94 | $4.64\times10^{32}$ | $4.27\times10^{33}$ |
| 170205 | 5445596501606963200 | 17.23 | 2.63 | 0.12 | 2.05 | 0.25 | 245.18 | 201.15 | $1.32\times10^{30}$ | 0.31 | $9.93\times10^{32}$ | $9.14\times10^{33}$ |
| 180609 | 5201723441529787392 | 19.12 | 2.97 | 0.19 | 2.36 | 1.21 | 682.82 | 122.25 | $8.58\times10^{29}$ | 8.81 | $5.16\times10^{33}$ | $4.75\times10^{34}$ |
| 180609 | 5201816139806400896 | 18.57 | 3.00 | 0.20 | 2.49 | 0.41 | 294.79 | 94.21 | $8.97\times10^{29}$ | 0.89 | $2.36\times10^{32}$ | $2.17\times10^{33}$ |
| 180609 | 5195663753775359232 | 20.66 | 3.39 | 0.19 | 2.86 | 1.56 | 146.53 | 111.87 | $1.74\times10^{29}$ | 4.09 | $1.04\times10^{32}$ | $9.60\times10^{32}$ |
| 180607 | 5201720555311609344 | 18.5 | 2.70 | 0.15 | 2.08 | 0.5 | 2051.84+ | 286.96 | $8.37\times10^{30}$ | 5.26 | $9.03\times10^{34}+$ | $8.30\times10^{35}$ |
| 180607 | 5190564291860908672 | 16.2 | 2.37 | 0.11 | 1.74 | 0.07 | 384.36 | 164.45 | $2.29\times10^{31}$ | 0.16 | $1.43\times10^{33}$ | $1.32\times10^{34}$ |
| 160730 | 5190981075487146368 | 18.5 | 2.32 | 0.08 | 1.64 | 0.11 | 289.53 | 356.78 | $1.29\times10^{31}$ | 0.38 | $1.44\times10^{33}$ | $1.33\times10^{34}$ |
| 160730 | 5191153870611593600 | 19.14 | 3.01 | 0.20 | 2.54 | 1.17 | 1161.24 | 159.25 | $1.43\times10^{30}$ | 7.43 | $1.23\times10^{34}$ | $1.13\times10^{35}$ |
| 150114 | 5501797275898167936 | 19.78 | 3.17 | 0.23 | 2.67 | 0.83 | 432.12 | 203.98 | $1.30\times10^{30}$ | 3.34 | $1.88\times10^{33}$ | $1.73\times10^{34}$ |
| 150114 | 5502892870516254720 | 19.04 | 2.90 | 0.20 | 2.32 | 0.24 | 580.09 | 185.39 | $2.12\times10^{30}$ | 1.64 | $2.02\times10^{33}$ | $1.86\times10^{34}$ |
| 150114 | 5502128435056792448 | 18.62 | 2.67 | 0.15 | 2.11 | 0.11 | 190.8 | 282.69 | $7.27\times10^{30}$ | 0.27 | $3.85\times10^{32}$ | $3.54\times10^{33}$ |
| 170205 | 5502472749699352320 | 17.17 | 2.82 | 0.17 | 2.28 | 0.14 | 243.50 | 75.21 | $2.08\times10^{30}$ | 0.31 | $1.56\times10^{32}$ | $1.44\times10^{33}$ |
| 170205 | 5502028757456041600 | 16.46 | 2.70 | 0.10 | 2.14 | 0.17 | 440.71 | 119.97 | $9.58\times10^{30}$ | 0.32 | $1.35\times10^{33}$ | $1.24\times10^{34}$ |
| 170206 | 5501843386667424640 | 20.16 | 3.39 | 0.23 | 2.99 | 1.07 | 242.77+ | 246.33 | $1.34\times10^{30}$ | 3.77 | $1.22\times10^{33}+$ | $1.13\times10^{34}$ |
| 170202 | 5502529344483998464 | 17.1 | 2.89 | 0.19 | 2.33 | 0.63 | 926.92 | 119.44 | $5.26\times10^{30}$ | 1.74 | $8.47\times10^{33}$ | $7.80\times10^{34}$ |
| 150115 | 5502495805083748736 | 20.7 | 3.04 | 0.23 | 2.50 | 1.48 | 479.87 | 425.62 | $2.43\times10^{30}$ | 13.81 | $1.61\times10^{34}$ | $1.48\times10^{35}$ |
| 150117 | 5502133893958218752 | 20.85 | 2.94 | 0.25 | 2.64 | 1.41 | 185.27 | 202.93 | $4.18\times10^{29}$ | 3.98 | $3.55\times10^{32}$ | $3.26\times10^{33}$ |
| 160726 | 6438530639589294720 | 19.11 | 2.65 | 0.13 | 2.07 | 0.51 | 336.32 | 419.48 | $1.02\times10^{31}$ | 2.19 | $7.54\times10^{33}$ | $6.93\times10^{34}$ |
| 160726 | 6435578657026534272 | 18.98 | 2.58 | 0.17 | 2.03 | 0.33 | 1059.51 | 343.08 | $7.69\times10^{30}$ | 3.24 | $2.64\times10^{34}$ | $2.43\times10^{35}$ |
| 160729 | 6435523131689270656 | 19.08 | 2.82 | 0.18 | 2.29 | 0.26 | 145.21 | 241.59 | $3.48\times10^{30}$ | 0.51 | $2.59\times10^{32}$ | $2.38\times10^{33}$ |
| 160729 | 6438148696738890880 | 17.19 | 2.74 | 0.18 | 2.19 | 0.07 | 385.34 | 107.44 | $3.92\times10^{30}$ | 0.26 | $3.94\times10^{32}$ | $3.64\times10^{33}$ |
| 160802 | 6435368817806876416 | 18.96 | 2.05 | 0.15 | 1.90 | 0.33 | 386.79 | 411.33 | $1.13\times10^{31}$ | 1.00 | $4.39\times10^{33}$ | $4.04\times10^{34}$ |
| 160802 | 6435441866611271296 | 20.14 | 2.89 | 0.17 | 2.39 | 1.73 | 288.99+ | 277.84 | $1.73\times10^{30}$ | 12.11 | $6.06\times10^{33}+$ | $5.58\times10^{34}$ |

| 160802 | 6438494836741682048 | 18 | 2.75 | 0.13 | 2.17 | 0.24 | 866.65 | 171.51 | $4.74 \times 10^{30}$ | 1.26 | $5.16 \times 10^{33}$ | $4.76 \times 10^{34}$ |
| 160802 | 6438484912872858816 | 18.72 | 2.29 | 0.09 | 1.64 | 0.39 | 289.22 | 465.59 | $1.80 \times 10^{31}$ | 0.75 | $3.90 \times 10^{33}$ | $3.59 \times 10^{34}$ |
| 160804 | 6435481586470460544 | 19.34 | 2.84 | 0.20 | 2.32 | 0.39 | 144.06 | 374.56 | $6.58 \times 10^{30}$ | 0.59 | $5.56 \times 10^{32}$ | $5.19 \times 10^{33}$ |
| 160806 | 6438355434986673152 | 20.01 | 2.50 | 0.12 | 2.01 | 0.44 | 240.73 | 345.81 | $3.03 \times 10^{30}$ | 1.53 | $1.11 \times 10^{33}$ | $1.02 \times 10^{34}$ |
| 160806 | 6435480903573366784 | 16.56 | 2.56 | 0.12 | 1.94 | 0.1 | 194.28 | 73.28 | $3.26 \times 10^{30}$ | 0.14 | $8.45 \times 10^{31}$ | $7.86 \times 10^{32}$ |
| 160728 | 6438601004035363584 | 21.01 | 3.02 | 0.20 | 2.61 | 1.02 | 440.41 | 308.19 | $9.56 \times 10^{29}$ | 5.30 | $2.23 \times 10^{33}$ | $2.05 \times 10^{34}$ |
| 160726 | 6400072128030483200 | 16.9 | 1.89 | -0.01 | 1.08 | 0.05 | 143.18 | 331.64 | $4.88 \times 10^{31}$ | 0.08 | $5.80 \times 10^{32}$ | $5.36 \times 10^{33}$ |
| 160726 | 6400154934999897344 | 14.81 | 1.32 | -0.02 | 0.60 | 0.05 | 144.81 | 334.91 | $3.41 \times 10^{32}$ | 0.05 | $2.37 \times 10^{33}$ | $2.18 \times 10^{34}$ |
| 160726 | 6400560066378703360 | 18.64 | 2.61 | 0.17 | 2.09 | 0.24 | 242.39 | 246.30 | $5.42 \times 10^{30}$ | 0.76 | $1.00 \times 10^{33}$ | $9.22 \times 10^{33}$ |
| 150114 | 5481255306315173248 | 16.9 | 2.40 | 0.11 | 1.77 | 0.05 | 482.67 | 170.72 | $1.29 \times 10^{31}$ | 0.27 | $1.71 \times 10^{33}$ | $1.58 \times 10^{34}$ |
| 150114 | 5482101964628144896 | 17.56 | 3.00 | 0.20 | 2.50 | 0.11 | 192.42 | 131.39 | $4.17 \times 10^{39}$ | 0.21 | $1.65 \times 10^{32}$ | $1.52 \times 10^{33}$ |
| 150116 | 5482088598688453504 | 22 | 1.80 | - | - | 2.25 | 1921.56 | 155.11 | $9.73 \times 10^{28}$ | 20.13 | $3.77 \times 10^{33}$ | $3.46 \times 10^{34}$ |
| 170202 | 4758254255601790592 | 18.71 | 2.53 | 0.08 | 1.94 | 0.63 | 664.45 | 353.73 | $1.05 \times 10^{31}$ | 2.32 | $1.62 \times 10^{34}$ | $1.49 \times 10^{35}$ |
| 170207 | 5482097188624710400 | 14.24 | 1.19 | -0.06 | 0.43 | 0.16 | 146.64 | 306.72 | $4.48 \times 10^{32}$ | 0.17 | $1.20 \times 10^{34}$ | $1.10 \times 10^{35}$ |
| 170202 | 4758720787833766400 | 19.89 | 2.38 | 0.10 | 1.67 | 0.19 | 237.89 | 427.90 | $5.17 \times 10^{30}$ | 0.40 | $4.94 \times 10^{32}$ | $4.54 \times 10^{33}$ |
| 170205 | 5482097115608455296 | 18.26 | 2.28 | 0.12 | 1.61 | 0.08 | 145.82 | 484.05 | $2.97 \times 10^{31}$ | 0.14 | $6.12 \times 10^{32}$ | $5.63 \times 10^{33}$ |
| 150116 | 4758312594141537152 | 20.65 | 3.33 | 0.24 | 2.71 | 1.08 | 1696.17 | 162.21 | $3.69 \times 10^{29}$ | 16.28 | $1.02 \times 10^{34}$ | $9.38 \times 10^{34}$ |

Table A1: DWF 500 pc flare sample properties.

[a] + indicates that the Flare fall was not fully within the time series data and this is a lower limit on the flare's duration.

[b] + indicates that the Flare fall was not fully within the time series data and this is a lower limit on the flare energy released.

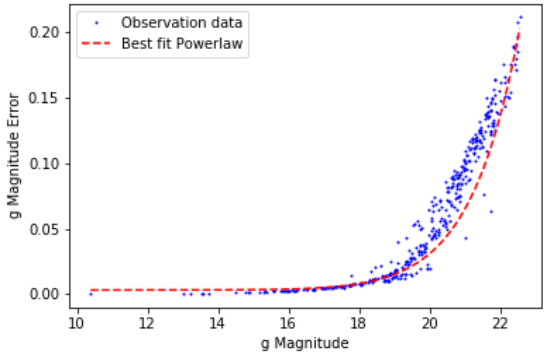

**Figure B1.** g magnitude error vs g magnitude for one central DECam CCD exposure of typical 20 s observations (e.g., airmass 1.5, seeing ~1″.0 FWHM).

## APPENDIX B:

This paper has been typeset from a TEX/LATEX file prepared by the author.